\documentclass[journal]{IEEEtran}
\usepackage[utf8]{inputenc}
\usepackage[T1]{fontenc}
\usepackage{float}
\usepackage{graphicx}
\usepackage{subfigure}
\usepackage{textcomp}
\graphicspath{{C:/Users/Dell/Desktop/LATEX/}}
\usepackage[export]{adjustbox}
\usepackage[table]{xcolor}
\usepackage{amsmath}
\usepackage{amsbsy}
\usepackage{multirow}
\usepackage{multicol}
\usepackage{url}

\begin{document}
\title{In-memory Implementation of On-chip Trainable and Scalable ANN for AI/ML Applications}
\author{Abhash Kumar,
        Jawar Singh,
        Sai Manohar Beeraka,
        and~Bharat Gupta% <-this % stops a space
\thanks{A Kumar, S M Beeraka, and B Gupta are with the Department
of Electronics and Communication Engineering, National Institute of Technology, Patna, Bihar, INDIA. abhash2205@gmail.com}% <-this % stops a space
\thanks{J Singh is with Electrical Engineering Department, Indian Institute of Technology Patna, Bihar, INDIA. dr.jawar@gmail.com }\vspace{-10mm} }
\maketitle

\begin{abstract}
Traditional von Neumann architecture based processors become inefficient in terms of energy and throughput as they involve separate processing and memory units, also known as~\textit{memory wall}. The memory wall problem is further exacerbated when massive parallelism and frequent data movement are required between processing and memory units for real-time implementation of artificial neural network (ANN) that enables many intelligent applications. One of the most promising approach to address the memory wall problem is to carry out computations inside the memory core itself that enhances the memory bandwidth and energy efficiency for extensive computations. This paper presents an in-memory computing architecture for ANN enabling artificial intelligence (AI) and machine learning (ML) applications. The proposed architecture utilizes deep in-memory architecture based on standard six transistor (6T) static random access memory (SRAM) core for the implementation of a multi-layered perceptron. Our novel on-chip training and inference in-memory architecture reduces energy cost and enhances throughput by simultaneously accessing the multiple rows of SRAM array per precharge cycle and eliminating the frequent access of data. The proposed architecture realizes backpropagation which is the keystone during the network training using newly proposed different building blocks such as weight updation, analog multiplication, error calculation, signed analog to digital conversion, and other necessary signal control units. The proposed architecture was trained and tested on the IRIS dataset which exhibits $\approx46\times$ more energy efficient per MAC (multiply and accumulate) operation compared to earlier classifiers.
\end{abstract}

 \vspace{-5mm}
\section{Introduction}
Artificial intelligence (AI) and machine learning (ML) algorithms are ubiquitous and integral part of contemporary computing devices, and significantly changing the way we live and interact with the world around us. Most of these computing systems are based on von Neumann architecture that involves separate processing and memory units where data need to be shuttled back and forth frequently between the processing and the memory units\cite{5389383}. Therefore, significant amount of the energy and time are consumed during data movement rather than actual computing, and this problem further exacerbated due to massive parallelism and data centric tasks such as decision making, object recognition, speech and video processing. This calls for a radical departure from the orthodox von Neumann approach to an unorthodox non-von Neumann computing architectures to carry out computations within the memory core itself, referred as to in-memory computing. Recently, hardware implementation of AI/ML algorithms based on in-memory computing has attracted huge attention because of unmatched computing performance and energy efficiency. In-memory computation overcomes the problem of frequent data movement between processing and memory units in the traditional von Neumann architecture based processors by carrying out computations within the memory core using its periphery circuitry.

%Ever since the introduction of the emerging fields of artificial intelligence, internet of things, etc., there have been an increasing demand of advanced algorithms related to these fields that are also data hungry. Further, most of the recent algorithms are software based and poses energy and delay optimization challenges to its hardware implementation due to large amount of data and computation required. In addition to this, most of the recent day processors are based on von-Neumann architecture which uses different memory and computing cores. Running data intensive algorithms, such as those related to artificial intelligence, neural networks, data analytics, etc., on such architecture is limited by \textit{von-Neumann bottleneck}\cite{5389383}. This bottleneck occurs due to limited throughput of the databus connecting the memory and computing core. Moreover, such data intensive algorithms will also lead to huge energy dissipation due to frequent transfer of huge data between memory and computing core. To address such issues many attempts have been done earlier out of which the in-memory computation is one of most the promising approach. In-memory computations overcomes the bottleneck problem in traditional von-Neumann based processors by carrying out computations right inside the memory core where the data is stored.\par 

Artificial Neural Network (ANN) is one of the most widely used tool for AI and ML based applications due to its very good resemblance and mimicking properties of human brain. It is used in a wide variety of AI/ML applications because of its self-learning ability to produce output that is not just limited to the input provided to them. Most of these AI/ML algorithms are software based and poses energy and delay optimization challenges for real time hardware implementation. The major limitations for hardware based solutions are large number of interconnections, massive parallelism, and time consuming calculations that requires huge data for training the networks and related algorithms. So, in-memory computation for such networks is one of the most preferred and efficient way for hardware realization of such complex networks. Researchers have come up with different in-memory implementations of popular machine learning networks and algorithms such as Convolution Neural Networks (CNNs)\cite{8345293, 7870353, 7573525}, Deep Neural Networks (DNNs)\cite{7870352, 7870351} and machine learning classifiers \cite{8094576, 7875410}. A large number of machine learning network architecture uses modified form of artificial neural networks such as in recurrent neural network (RNN), the fully connected layer of CNN, or in Deep Neural Networks (DNNs) which are the foundations of deep learning. There is a remarkable improvement in learning and predicting the complex pattern of a large data set which otherwise would have been very difficult even for the Graphic Processing Units (GPUs) that still require on-chip or off-chip memory access despite parallel computations for inference and training of these algorithms. 

%Neural Network is one of the most widely networks for artificial intelligence applications due to the way it mimics the behaviour of human brain. It is used in a wide variety of artificial intelligence applications because of its self-learning ability and produces output that is not just limited to the input provided to them. But due to a large number of interconnections, complex and time-consuming calculations, huge data required for training, the hardware implementation of these networks, and related algorithms are very difficult to achieve. So, in-memory modelling of such networks is one of the most preferred way for hardware realization of such complex networks. Researchers have come up with such in-memory implementations of popular machine learning networks and algorithms such as Convolutional Neural Networks (CNN)\cite{8345293,7870353,7573525}, Deep Neural Network (DNN)\cite{7870352,7870351} and machine learning classifiers \cite{8094576,7875410}. A large number of machine learning network architecture uses neural network, such as the fully connected layer of CNN, or being the backbone of Deep Neural Networks it gives the state of art results in learning and predicting complex pattern of a large dataset which otherwise would have been very difficult even for the machines to perform using normal straight forward algorithms.\par 

Most of the AI/ML algorithms require processing of a large data sets and the energy cost involved is mostly governed by the frequent access of memory\cite{6757323}, especially for dense networks such as DNNs\cite{7738524, 10.1145/2654822.2541967}. In DianNao\cite{10.1145/2654822.2541967} and Eyeriss\cite{7738524}, the data reuse have been an efficient and highly effective solutions for saving energy, however, it still resulted in frequent on-chip memory access leading to 35\% to 45\% of total energy dissipation. Other techniques, such as reducing the precision of parameters to even 1-b during inference\cite{courbariaux2016binarized, rastegari2016xnornet, 8465935} can address the energy and latency issues to a some extent but they will lead to accuracy trade-off. Further, implementation of architectures in digital domain using low power circuits have been employed such as power gating technique for speech recognition\cite{7870352}, RAZOR\cite{1253179} for Internet of Things (IoT) applications\cite{7870351}, and architectural designs such as dynamic-voltage accuracy frequency scalable CNN processor\cite{7870353} have been introduced earlier to reduce energy cost. They exploits the advantages of scalability and programmability of implementations in digital domain but they miss out the opportunities that lies in analog domain for implementation of AI/ML algorithms. This is due to the unique data flow of ANN during feedforward and backpropagation that can be exploited to design energy efficient and high throughput in-memory ANN architectures in analog domain. 

Exploiting the opportunities available in analog domain for realizing in-memory AI/ML algorithms to reduce energy and delay cost, an in-memory inference processor based on \textit{deep in-memory architecture} (DIMA)\cite{8246704,compute} have been presented earlier. The DIMA stores binary data in a column-major format as opposed to the row-major format employed in traditional Static Random Access Memory (SRAM) array organization. It reduces energy cost by simultaneously accessing multiple rows of the standard six-transistor (6T) SRAM cells per precharge cycle through the application of modulated pulse width signals to the word lines (WLs), and thus increases the throughput. Previously, DIMA is also used for AI/ML algorithms such as CNNs\cite{8345293} and its versatility was also established by mapping the ML algorithms for template matching\cite{6855225}, and architectures of sparse distributed memory\cite{7169194, 7489032}. A multi-functional in-memory inference processor\cite{8246704} based on DIMA have been presented earlier which achieves $53\times$ reduction in energy delay product (EDP) and supports four algorithms: support vector machine, \textit{k}-nearest neighbour, template matching, and matched filter. But, the biggest disadvantage of DIMA was that it cannot be used for supervised learning algorithms which requires training of the network. This is because the underlying hardware of the DIMA does not support on-chip training of the network. Recent work on DNN\cite{9035482} deals with weight update but it requires additional buffers for storing the output before performing convolution at each stage which increases energy cost involved in frequent storage and access of data from these buffers which also increases latency.

In this paper, a DIMA based memory core is employed with proposed peripheral circuitry to realize in-memory ANN architecture. The memory read cost of the proposed architecture is reduced via DIMA functional read (DIMA-FR) process which access the multiple rows of SRAM bit cell arrays (BCA) per precharge cycle. Most of the hardware based AI/ML architectures do not support on-chip training and even if some of them does, then they do not able to avoid the frequent access to memory core during each iteration of training contributing to a large proportion in total energy and delay. To alleviate the memory access bottleneck, sampling capacitors have been used in the proposed architecture that store the weights temporarily and avoid the frequent memory access during each iteration. The proposed architecture provides on-chip hardware support for the feedforward (FF), backpropagation (BP), error calculation as well as weight updation. Further, it supports multilayered and scalable architecture by cascading the proposed single layered neural network. Each of these layer communicates with its preceding and next layer in the same way a neural network does. Moreover, a large variety of AI/ML algorithms can be implemented using the proposed architecture by just changing the activation function block and the number of memory banks in the architecture according to the requirements. Overall, the proposed architecture exactly mimics the way a neural network works and it is configurable to realize variety of neural networks based AI/ML algorithms. Simulation results demonstrate that the proposed architecture achieves \(\approx 1.04\times\) reduction in energy delay product (EDP) and \(\approx 46\times\) reduction is energy per mutiply-and-accumulate (MAC) operation. In contrast to existing in-memory computing architectures, following are the key contributions of the proposed architecture:
\begin{enumerate}
    \item \textbf{Hardware support for on-chip training--} Hardware support for feedforward and backpropagation process is designed in the periphery of SRAM bitcell array that mimics the way a neural network works. Further, the weight update mechanism has been developed without the need of intermediate buffers to avoid latency and energy cost involved in read/write mechanism of such buffers. Moreover, appropriate control signals have been designed for parallelizing weight update process of the whole network.
    \item \textbf{On-chip error calculation--} An on-chip sum of squares of error calculating mechanism have been developed whose gradients with respect to weights are back propagated for network training.
    \item \textbf{Signed flash ADC design--} The signed flash ADC is designed and presented to stored updated signed weights back inside SRAM memory core which follows \(1's\) complement conversion method for negative weights.
\end{enumerate}

\begin{figure*}[ht]
    \centering
    \includegraphics[width=170mm]{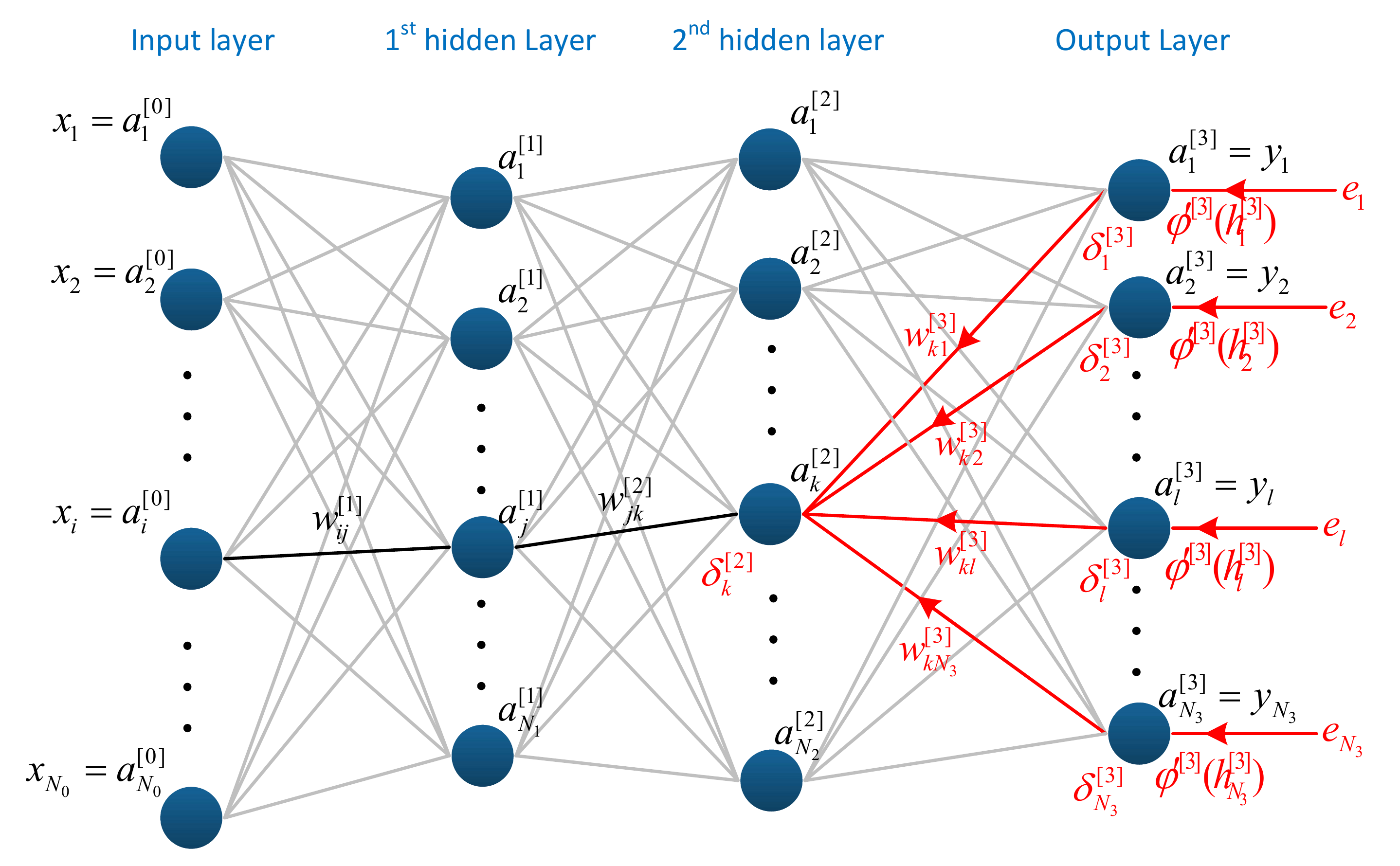}
    \caption{Network showing multilayered perceptron with two hidden layers. The signals in black are during feedforward process and the signals in red shows the backpropagation from the output layer to the \(k^{th}\) neuron of the \(2^{nd}\) hidden layer.}
    \label{fig_Multilayered_Perceptron}
    \vspace{-5mm}
\end{figure*}
\vspace{-3mm}

\section{Background}
This Section discusses the pre-requisites for the development of artificial neural network\cite{haykin2010neural} and hardware implementation of deep in-memory architecture (DIMA)\cite{8246704,compute}. \vspace{-3mm}
\subsection{Neural Network}
The neural network is a complex network of neurons where each of the connections is characterized by synaptic weights. There are three basic elements of neuron:\par(1) \textit{set of synapses or connecting links}-- each characterized by a weight, \par (2) \textit{an adder}-- for summing the products of input signal and corresponding synapse weight, and \par (3) \textit{an activation function}-- typically sigmoid, tanh, or ReLU that limits the output of a neuron. 
Fig.~\ref{fig_Multilayered_Perceptron} shows the generalized multilayered perceptron neural network exhibiting the signals at each layer during feedforward (in black colour) and backpropagation (in red colour). Both feedforward and backpropagation have been shown in the same network for a better understanding of signal flow during each of these two processes. In Fig.~\ref{fig_Multilayered_Perceptron}, \(N_K\) (for \(K\in W\), where \(W\) denotes the whole number) corresponds to total number of neurons in the \(K^{th}\) layer (\(K=0\) corresponds to the input layer). The \(w_{ij}^{[K]}\) denotes weight of the synaptic connection to the \(j^{th}\) neuron (of the \(K^{th}\) layer) from the \(i^{th}\) neuron (of its previous layer). In this paper, the subscripts \(i, j, k,\) and \( l\) (for \(i, j, k, l\in W\)) are used for indexing the properties/signals associated with a particular neuron and the superscript \([K]\) for indicating the layer to which it is associated with. Paired subscript, such as \textquoteleft\(ij\)\textquoteright shows that the property is associated with the connection of \(i^{th}\) neuron of any layer to the \(j^{th}\) neuron of its next layer. The training of the network takes place in two phases using the backpropagation algorithm\cite{haykin2010neural}.  

\subsubsection*{Forward Phase}
In forward phase, the weights of the network are initialized and the input signals are propagated layer by layer through the entire network until they reach to the output layer. From Fig.~\ref{fig_Multilayered_Perceptron}, the output at the \(K^{th}\) layer is given by the following matrix multiplication and subsequently by the application of activation function on it as shown below:
    \begin{equation}
    \label{eqn_Output_of_Neuron}
    \begin{bmatrix}
        a_1^{[K]}\\a_2^{[K]}\\\vdots\\a_{N_K}^{[K]}
    \end{bmatrix}
    =\varphi\left(
    \begin{bmatrix}
    w_{11}^{[K]} & w_{21}^{[K]} & ... & w_{N_{K-1}1}^{[K]}\\ 
    w_{12}^{[K]} & w_{22}^{[K]} & ... & w_{N_{K-1}2}^{[K]}\\
    \vdots & \vdots & . & \vdots\\ 
    w_{1N_{K}}^{[K]} & w_{2N_{K}}^{[K]} & ... & w_{N_{K-1}N_{K}}^{[K]}\\
    \end{bmatrix}
    \begin{bmatrix}
        a_1^{[K-1]}\\a_2^{[K-1]}\\\vdots\\a_{N_K}^{[K-1]}
    \end{bmatrix}
    \right)
    \end{equation}
    where \(\varphi(\cdot)\) is the activation function at the output of the neurons in the \(K^{th}\) layer.
\subsubsection*{Backward Phase}
Once the final output is available during the forward phase, then the error is calculated by comparing the final output with the target (or intended) output. The error at any output neuron is calculated as \(e_l=t_l-y_l\), where \(y_l\) is the current output and \(t_l\) is the target (or intended) output at the \(l^{th}\) output neuron. For calculating the total error of the network there are many cost functions available of which the sum of squares of error is the most popular one which is given as:
\begin{equation} \label{eqn_square of error}
E=\frac{1}{2}\sum_{l=1}^{N_3}e_l^2
\end{equation}
In Eq.~\ref{eqn_square of error}, the square of error will be summed up for all the output neurons in Fig.~\ref{fig_Multilayered_Perceptron}. The resulting error is propagated through the network in the backward direction and successive weight adjustments/updates are made to the synaptic weights using gradient descent algorithm. The gradient descent algorithm states that weight should be moved in the direction of negative gradient of the error landscape to minimize the error of the network. Mathematically, the weight update of the synaptic connections to the \(K^{th}\) layer is given as:
\begin{equation} \label{eqn_weight_update}
    w_{jk}^{[K]}\leftarrow w_{jk}^{[K]}-\eta\frac{\partial E}{\partial w_{jk}^{[K]}}=w_{jk}^{[K]}+\Delta w_{jk}^{[K]}
\end{equation}
where \(E\) is the error as calculated in Eq.~\ref{eqn_square of error} and \(\Delta w_{jk}^{[K]}\left(=-\eta\partial E/\partial w_{jk}^{[K]}\right)\) is the change in weight required to reduce the error of the network. Now, depending upon where the neuron is situated in the network, two cases arise. Firstly, when the neuron will be at the output layer of ANN. In this case, the desired response at each of the output node is supplied during training, making this case a straightforward to handle from error calculation and weight updation point of view. Secondly, when the neuron is situated at any hidden layer. Hidden layer neurons are not directly accessible but they do share the responsibility for the total error of the network. At the same time their desired response is not known beforehand which makes the estimation of their contribution in total error even more difficult. Both cases are discussed in detail as follows:
\begin{itemize}
    \item \textit{\textbf{CASE I: \textquoteleft\(K\)\textquoteright denotes an output layer--}} The desired response at an output neuron is known in this case as discussed earlier, so Eq.~\ref{eqn_square of error} (and Fig.~\ref{fig_Multilayered_Perceptron}) can be used directly to calculate the weight change as:
    \begin{equation} \label{eqn_Weight_update_output}
        \Delta w_{kl}^{[3]}=\eta\left(t_l-y_l\right) \varphi '^{[3]}\left(h_l^{[3]}\right)a_k^{[2]}=\eta\delta_l^{[3]}a_k^{[2]}
    \end{equation}
    where \(h_l^{[3]}=\sum_{k=1}^{N_2}a_k^{[2]}w_{kl}^{[3]}\) is the activation potential at the \(l^{th}\) neuron of the output layer, \(\delta_l^{[3]}=(t_l-y_l)\varphi'^{[3]}\left(h_l^{[3]}\right)\) is the local gradient at \(l^{th}\) neuron of the output layer of Fig.~\ref{fig_Multilayered_Perceptron}, and \(\varphi'^{[3]}(\cdot)\) is the derivative of the activation function.
    \item \textit{\textbf{CASE II: \textquoteleft\(K\)\textquoteright denotes a hidden layer--}} There is no specified desired response in this case, hence the error signal has to be determined recursively and working backward in terms of the error signals of all the neurons to which that hidden neuron is directly connected. Using Eq.~\ref{eqn_square of error}, and applying chain rule the simplified result\cite{haykin2010neural} can be obtained as:
    \begin{equation} \label{eqn_weight_update_hidden}
        \Delta w_{jk}^{[2]}=\eta\left(\sum_l^{N_3}\delta_l^{[3]}w_{kl}^{[3]}\right)\varphi'^{[2]}\left(h_k^{[2]}\right)a_j^{[1]}=\eta\delta_k^{[2]}a_j^{[1]}
    \end{equation}
    where \(\delta_k^{[2]}=\sum_{l=1}^{N_3}\left(\delta_l^{[3]}w_{kl}^{[3]}\right)\varphi'^{[2]}\left(h_k^{[2]}\right)\) is the local gradient at the \(k^{th}\) neuron of the \(2^{nd}\) layer of the Fig.~\ref{fig_Multilayered_Perceptron}. Similarly, the weight update for the \(1^{st}\) layer is given as:
    \begin{equation} \label{eqn_weight_update_input}
       \Delta w_{jk}^{[1]}= \eta\delta_k^{[1]}a_j^{[0]}
    \end{equation}
\end{itemize}
This is how error is propagated backward in the network. This back propagation of error is shown in red color in Fig.~\ref{fig_Multilayered_Perceptron}. The above derivations are described for a multilayered perceptron neural network with only two hidden layers, however, the same concept can be extended to any number of hidden layers.\par

\begin{figure*}[t]
    \centering
    \includegraphics[width=175mm]{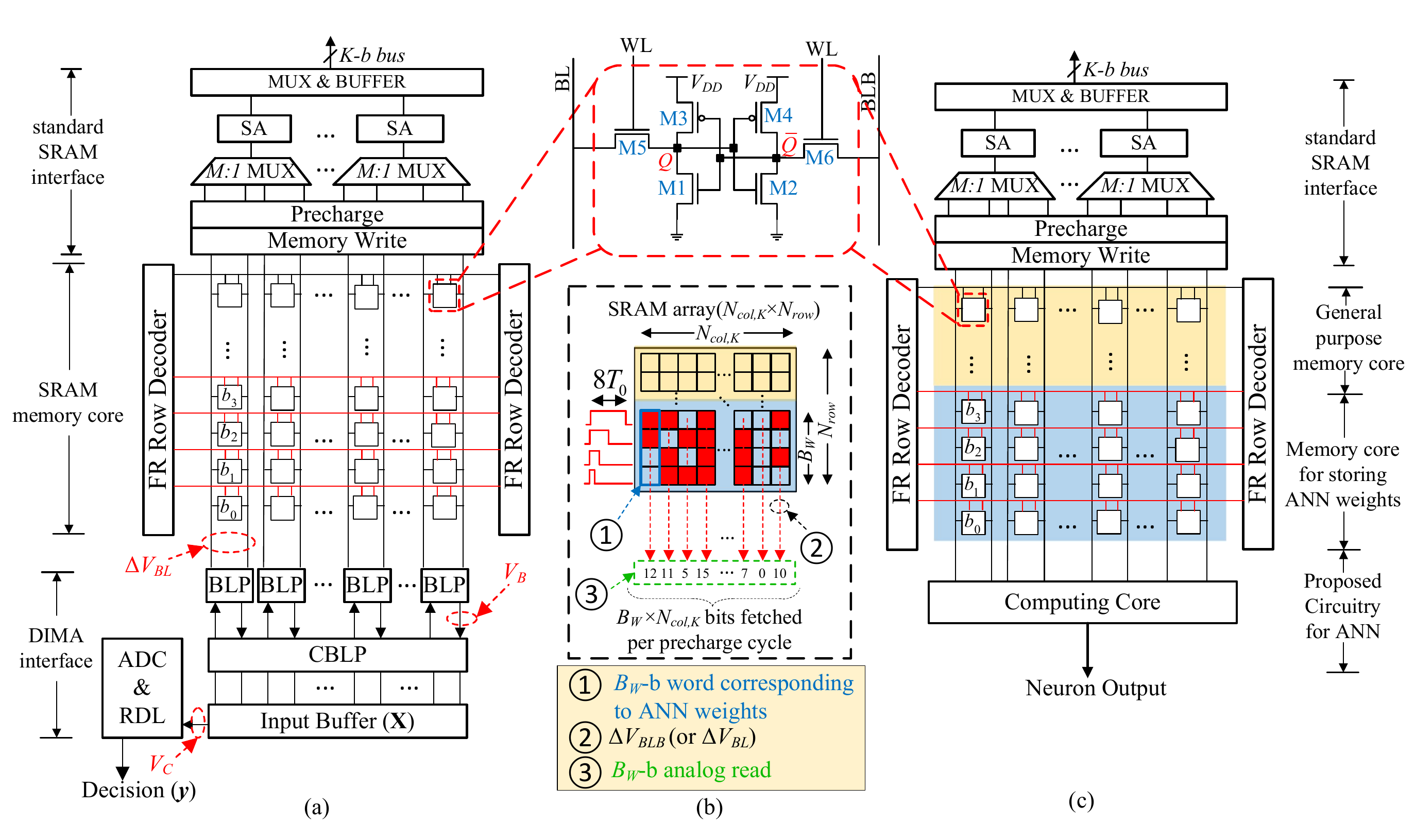}
    \caption{(a) Conventional deep in-memory architecture (DIMA)\cite{8246704}. (b) Memory access via functional read (FR) process of the DIMA for four bits per weight (\(B_W=4\)). (c) Generalized block diagram of the proposed architecture which employs FR of DIMA for accessing weights stored in last \(B_W\) rows of SRAM bitcell array (BCA).}
    \label{fig_Single_Bank_DIMA} \vspace{-3mm}
\end{figure*}
\vspace{-3mm}
\subsection{Basics of In-Memory Architecture}
The main and mature element of in-memory computing is the standard six transistor (6T) static random access memory (SRAM) cell/array, where most of the computations are performed in parallel and mixed-signal domain  which results in significant improvement in throughput and energy efficiency. Recently, a 6T-SRAM based deep in-memory architecture (DIMA)\cite{8246704} has demonstrated very good energy efficiency for many intelligent and data intensive applications. Fig.~\ref{fig_Single_Bank_DIMA}(a) shows the conventional DIMA architecture. The DIMA uses the standard SRAM bit cell array (BCA) with read/write circuitry at the bottom and stores \(B_W\) bits of weight in a column-major format as opposed to the row-major format used in conventional SRAM cell. The DIMA consists of four processes that are executed sequentially. \par
(1)~\textit{Multi-row Functional READ (FR)}-- that performs digital to analog conversion of weights \(w\) (index \(i,j,k,l\) and \([K]\) are omitted for simplicity) stored in standard 6T SRAM cell by fetching \(B_W\) bits of weight which is achieved by simultaneously reading \(B_W\) rows per pre-charge cycle, \par
(2)~\textit{ Bit Line Processing (BLP)}-- calculates scalar distances (SDs) by carrying out mathematical operations (such as - addition, subtraction, multiplication, etc.) between weights stored in SRAM cells and the applied input signal,\par
(3)~\textit{ Cross BLP (CBLP)} -- carries out the summation of scalar distances (SDs) via charge sharing across the required columns of the bit cell array to calculate the vector distances (VDs) such as dot product if SD is multiplication in BLP stage, or Manhattan distance if SD is absolute difference in BLP stage, and \par
(4)~\textit{Analog-to-Digital Converter (ADC) and Residual Digital Logic (RDL)}-- stage for realizing thresholding/decision function and carrying out analog to digital conversion of the results of previous analog computations.\par

Fig.~\ref{fig_Single_Bank_DIMA}(b) shows the multi-row functional read process of the DIMA. The multi-row functional read stage yields a voltage discharge \(\Delta V_{BL}\) proportional to the weighted sum \(\overline{w}=\sum_{i=0}^{B_W-1}2^i\overline{b_i}\) of column-major stored digital data \(\left\{b_0, b_1,..., b_{B_W-1}\right\}\) by applying binary-weighted modulated pulse width signal \(T_i\propto2^i\left(i\in\left[0,B_W-1\right]\right)\) to \(B_W\) rows of SRAM array simultaneously, as shown in Fig.~\ref{fig_Single_Bank_DIMA}(b). The voltage drop on bitline (BL) as a function of weight \(w\) is given by:
\begin{equation} \label{eqn_BL_line_discharge}
    \Delta V_{BL}=\frac{V_{PRE}}{R_{BL}C_{BL}}T_0\sum_{i=0}^{B_W-1}2^i\overline{b_i}=\Delta V_{lsb}\overline{w}
\end{equation}
where \(\Delta V_{lsb}=\frac{V_{PRE}}{R_{BL}C_{BL}}T_0\), \(V_{PRE}\) is the precharge voltage on BL and its complement (BLB) lines, \(R_{BL}\) and \(C_{BL}\) are the discharge path resistance and capacitance respectively, \(\overline{w}\) is the decimal equivalent of \(1's\) complement of weight stored in SRAM cell array in column-major format. Similarly, the voltage drop at the complement of BL (i.e. BLB) line can be expressed by simply replacing \(\overline{w}\) with \(w\) in Eq.~\ref{eqn_BL_line_discharge}, i.e.,:
\begin{equation} \label{eqn_BLB_line_discharge}
    \Delta V_{BLB}=\frac{V_{PRE}}{R_{BL}C_{BL}}T_0\sum_{i=0}^{B_W-1}2^ib_i=\Delta V_{lsb}w
\end{equation}
It is important to note that both Eqs.~\ref{eqn_BL_line_discharge} and ~\ref{eqn_BLB_line_discharge} are valid for \(T_i\ll R_{BL}C_{BL}\). A sub-ranged read technique \cite{8246704} can be employed for improving the linearity of the FR process when \(B_W>4\) for which \(B_W/2\) bits representing the most significant bits (MSB) and \(B_W/2\) bits representing the least significant bits (LSB) of the weights stored in adjacent columns of the BCA. For this, the MSB and LSB BL capacitances \(C_M\) and \(C_L\), respectively, are required to be in the ratio \(2^{B_W/2}:1\). The MSB and LSB columns are first read separately and then merged by assigning more weights to MSB than LSB using ratioed capacitors \(C_M\) and \(C_L\).

However, the inference processor presented in ref. \cite{8246704} supports four algorithms - Support Vector Machine, Template Matching, \textit{k}-Nearest Neighbour, and Matched Filter which have been just mapped to them without discussing the way to train the network on the chip. Further, the peripheral circuits used around memory core in standard DIMA architecture lacks the hardware support for backpropagation algorithm and weight update mechanism which is a crucial part in training any neural networks. Also, frequent access to SRAM bitcell array during calculation of scalar distance (SD) at each iteration in conventional DIMA\cite{8246704} architecture leads to a significant amount energy dissipation and latency. To realize hardware implementation of ANN along with addressing above discussed issues related to DIMA and other architectures discussed in introductory part of this paper, an in-memory ANN architecture is proposed in next Section.

\begin{table}[t]
\centering
    \begin{tabular}{|l|l|}
    \hline \rowcolor{lightgray}
        Notation & Description\\\hline
        ADC & Analog to Digital Converter\\\hline
        ANN & Artificial Neural Network\\\hline
        BCA & Bit Cell Array\\\hline
        BL & Bit Line\\\hline
        BLB & Bit Line Bar\\\hline
        \(B_W\) & No. of bits per weight\\\hline
        DIMA & Deep In-Memory Architecture\\\hline
        FR & Functional Read\\\hline
        \(M=\left(N_{K-1}\right)\) & No. of inputs to the \(K^{th}\) layer\\\hline
        \(N=\left(N_K\right)\) & No. of output of the \(K^{th}\) layer\\\hline
        \(N_{bank,K}\) & No. of banks for the \(K^{th}\) layer\\\hline
        \(N_{col,K}\) & No. of columns per memory bank for the \(K^{th}\) layer \\\hline
        \(N_{row}\) & No. of rows in the bit cell array\\\hline
        SA & Sense Amplifier\\\hline
        SWC & Signed Weight Calculation\\\hline
        SM & Signed Multiplier\\\hline
        \(MT\) lines & \(M\)-transmission lines\\\hline
        \(NT\) lines & \(N\)-transmission lines\\\hline
        WL & Word Line\\\hline
        WU & Weight Updation\\
        \hline     
    \end{tabular}
    \caption{Acronyms and Notations}
    \label{Table_AcronymsAndNotations} \vspace{-2mm}
\end{table}
\vspace{-3mm}
\section{The Proposed In-Memory ANN Architecture}
In this Section, the proposed In-Memory Artificial Neural Network architecture is presented. First, detailed architecture and working of a single layer of a neural network is discussed and then it is extended to the multilayered perceptron consisting of interconnections of many such layers. Fig.~\ref{fig_Single_Bank_DIMA}(c) shows a generalized block diagram of single DIMA-based memory bank. It is a basic building block of our proposed in-memory ANN architecture. Each memory bank corresponds to a single output neuron. As discussed in previous Section, the peripheral BCA circuitry of the traditional DIMA architecture is incapable of carrying out on-chip training of the network, therefore, we have incorporated the additional peripheral circuitry that facilitate the on-chip training and supports both feedforward and backpropagation processes for realizing the in-memory ANN. The proposed in-memory ANN architecture utilizes the last \(B_W\) rows of the BCA for storing weights (in column-major format) of the synaptic connections of an ANN and are accessed via multi-row FR process which yields the analog equivalent of stored weight as shown in Fig.~\ref{fig_Single_Bank_DIMA}(b). The rest of the memory space, i.e., \(\left[N_{col,K}\times\left(N_{row}-B_W\right)\right]\) bits of the BCA can be used as a general purpose digital storage medium.

\begin{figure*}[t]
    \centering
    \includegraphics[width=175mm]{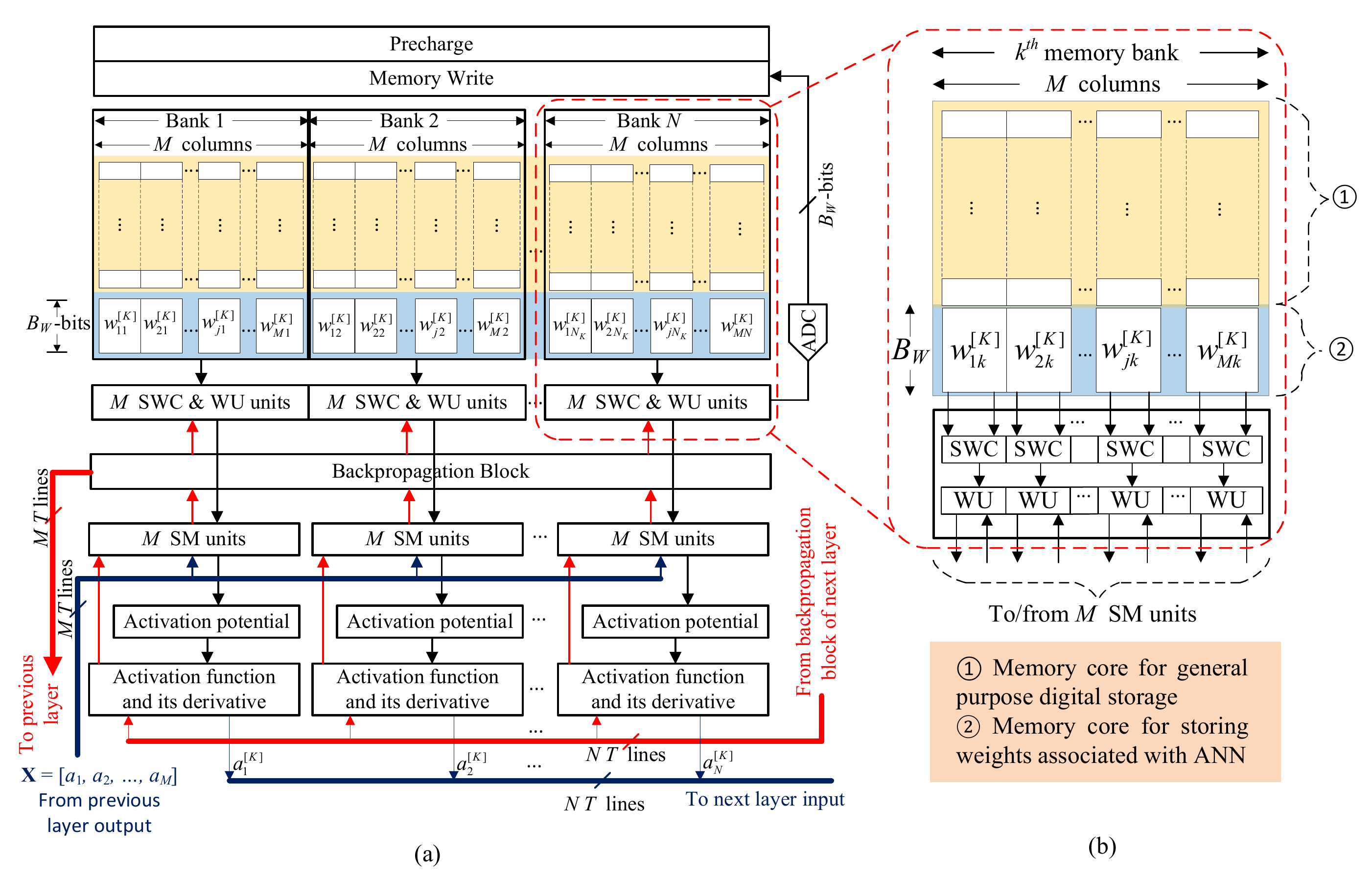}
    \caption{(a) Block diagram showing proposed in-memory multi-bank implementation of ANN. The T. lines (transmission lines) carrying forward propagating signals are shown in blue color; and the T. lines (transmission lines) carrying backpropagating signals are shown in red color. (b) A single DIMA based memory bank connected to \(M\) SWC \& WU units.}
    \label{fig_Single_Layer_Proposed} \vspace{-5mm}
\end{figure*}

Fig.~\ref{fig_Single_Layer_Proposed} shows the proposed in-memory ANN architecture of any arbitrary layer of ANN. We have used \(M\) and \(N\) as general terms indicating the number of inputs and outputs, respectively, of the \(K^{th}\) layer. From Fig.~\ref{fig_Single_Layer_Proposed}, multi-bank division of the BCA have been done for parallelizing the computations involved in each layer of the ANN. Each bank is used to store weights of the connections from the outputs of the previous layer to the input of a neuron in the present layer. Each bank consists of \(M\) columns, for storing weights of the connection from \(M\) inputs to an output of this layer. Further, the number of such banks, \(N_{bank,K}\), depends on the number of neurons at the output of this layer which is assumed to be \(N\) in our case. The FR process performs the analog to digital conversion of weights by discharging the BLB and BL lines by an amount proportional to their decimal equivalent \(\left(w\right)\) and its \(1's\) complement \(\left(\overline{w}\right)\), and can be derived from Eqs.~\ref{eqn_BLB_line_discharge} and ~\ref{eqn_BL_line_discharge}, respectively. Since, the weights can be either positive or negative, so a signed weight calculation (SWC) unit is designed for generating these weights in terms of proportional negative voltage (for negative weights) or positive voltage (for positive weights). The signed weights thus will be generated and sent to the weight updation (WU) unit where weights are updated in each iteration and then sent to the signed multiplier (SM) unit, these peripheral circuits are essential for on-chip training that makes our proposed approach energy efficient and real time. 

The input vector \(\textbf{X}=\left[a_1,a_2,...,a_M\right]\) (index \(i,j,k,l\) and \([K]\) have been omitted to avoid confusion and \(\left[a_1,a_2,...,a_M\right]\) is taken as the inputs to any general layer, i.e., the \(K^{th}\) layer) to this layer are fetched via \(M\)-transmission (\(M T\)) lines  each storing analog voltages proportional to the inputs to the layer, as shown in Fig.~\ref{fig_Single_Layer_Proposed}(a). These inputs are sent to the SM units where it is multiplied with signed weights, calculated via SWC units during feedforward process, to generate product of input and signed weight. To generate activation potential we need to sum these individual products at each of the output neuron, i.e, \(\sum_{j=1}^{M}a_j^{[K-1]}w_{jk}^{[K]}\) for which current based summation of weighted inputs is employed inside the activation potential block (discussed in upcoming sub-sections). The output of the activation potential block is sampled and stored in capacitor to avoid its regeneration during backpropagation where it will be used for weight update. This activation potential is fed to the activation function block from where the final output of the layer is obtained. The output of the layer are sent to the next layer through \(NT\) lines.

During backpropagation, the signals traverse through various components of the network and they will be governed by Eqs.~\ref{eqn_Weight_update_output}, ~\ref{eqn_weight_update_hidden} and ~\ref{eqn_weight_update_input}. The weighted sum of the local gradients (the weights applied to the local gradient are the weights of synaptic connections between the layer having the local gradients and its previous layer) of any layer are propagated backward to the previous layer which is multiplied with its input and learning rate for its weight update. Also, from Eqs.~\ref{eqn_Weight_update_output}, ~\ref{eqn_weight_update_hidden}, and ~\ref{eqn_weight_update_input}, for generating local gradient at any neuron in the present layer, the weighted summation of the local gradient coming from the next layer is multiplied with the derivative of the activation function of the present layer neuron. In our proposed architecture, the weighted sum of the local gradients of any layer, i.e., the \(K^{th}\) layer is generated via current based summation which is employed inside the backpropagation block and is propagated to the previous layer through \(MT\) lines coming out, as shown in Fig.~\ref{fig_Single_Layer_Proposed}(a)). For the weight update, the local gradient of this layer is multiplied with input to this layer and sent to the weight updation (WU) unit. The learning factor \(\eta\) can be controlled by controlling the gain at the output of the multiplier going to WU unit. During feedforward and backpropagation processes, different signals need to be multiplied together which is controlled by control signal \(S[1:0]\). Once the training of the network finishes, the WU unit will have all the trained weights stored in it. These trained weights need to be stored in BCA memory core since these weights are analog and susceptible to noise, weights may also drift due to noise/charge leakage then entire network will have to be trained again to avoid accuracy loss. To address weight drifting issue, signed flash ADC is designed and presented which will convert the signed weight to digital domain considering \(1's\) complement conversion for negative weights since the BL line have voltage discharge proportional to decimal equivalent of \(1's\) complement of weight which can be used to obtain absolute value of the negative weight during FR read process which can be further converted to proportional negative voltage. The output of signed flash ADC will be stored in SRAM BCA. The upcoming Subsections present detailed description of these blocks to support in-memory ANN essentials for on-chip training.
\vspace{-3mm}
\subsection{Signed Weight Calculation (SWC) Unit}
Fig.~\ref{fig_Single_Layer_Proposed}(b) shows a single bank DIMA with SWC and weight updation (WU) units for generating signed and updating the weights, respectively. The weights of synaptic connections can either be negative or positive which will require a hardware circuitry that must be equipped with a suitable scheme to realize signed dot product. The signed dot product calculator has been proposed earlier in ref.\cite{8345293} which has two rails: CBLP+ rail (for positive product) and CBLP– rail (for negative product) shared with multiple columns each corresponding to the output of a BLP block. The BLP block was used there to select the magnitude of weight and then calculate the product of input vector with weights stored in SRAM. These products were then shared with CBLP+ rail for positive weights and CBLP- rail for negative weights. These weights were accessed via DIMA FR process to get proportional analog equivalent of weight. These selection of BL or BLB lines for getting absolute value of weight and selection of CBLP+ or CBLP- rail depending on sign of product were done using control circuitry inside BLP block itself. These rails were then passed through the ADC and fed to the subtractor block which calculates the difference between them to get vector distance. However, this approach is inefficient for updating the weights during each iteration (on-chip training) because we are performing computations in analog domain to ease the process of weight updation by storing intermediate weights in capacitors (discussed in the next Sub-section). If computations were done in digital domain then it would have required additional registers/buffers to store intermediate weight during training consuming large silicon chip area for dense network involving a large number of weights. 

\begin{figure*}[t]
    \centering
    \includegraphics[width=168mm]{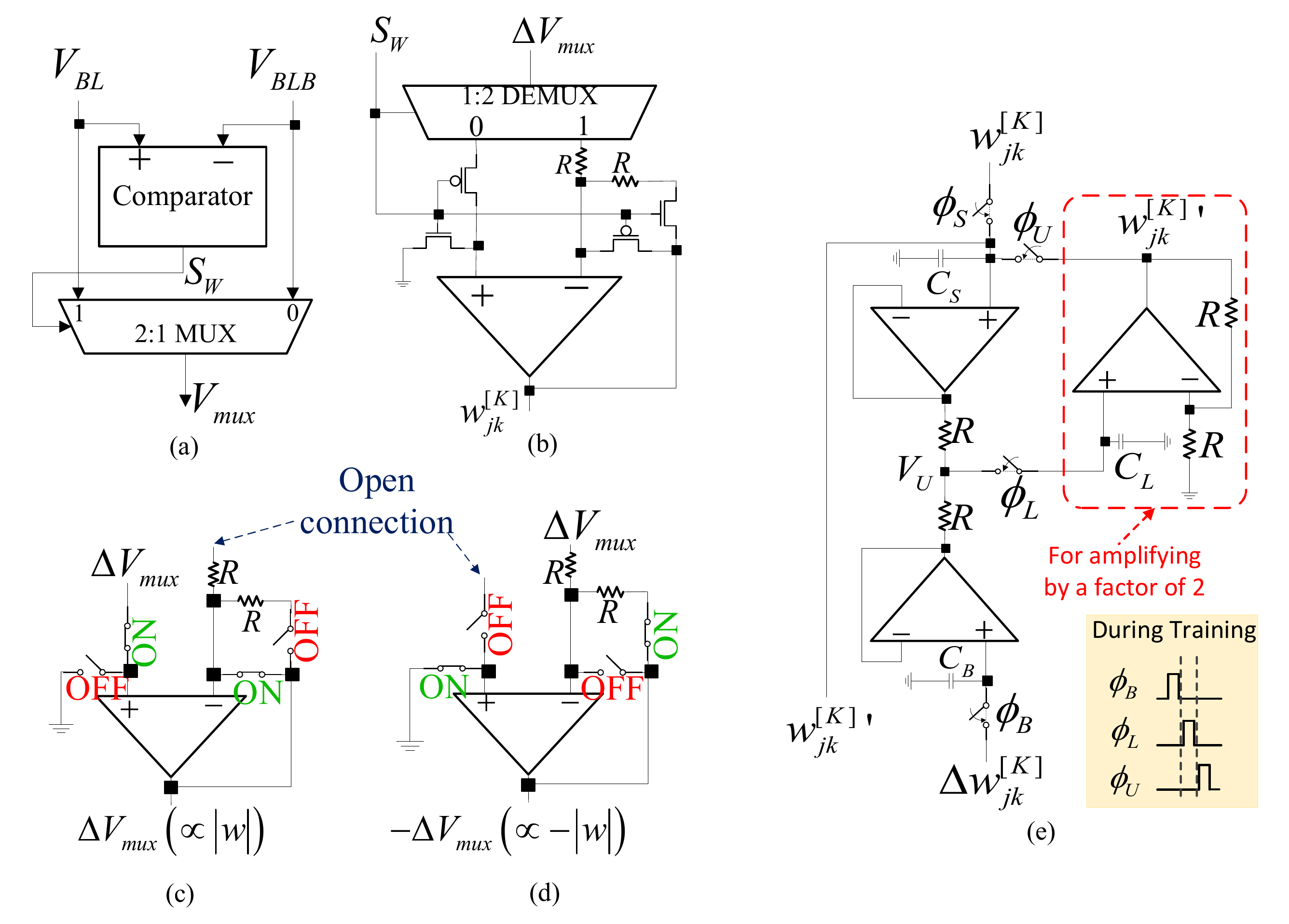}
    \caption{Proposed hardware design for signed weight calculator (SWC) and weight updator (WU) units. (a) Selecting bit line (BL or BLB) containing magnitude of weight. (b) sending the change in MUX voltage, i.e., \(\Delta V_{mux}\) to one of the inputs of OPAM to perform either (c) unity gain follower operation for positive weight, i.e., \(S_W=0\) or (d) inverting the polarity of weight for negative weight, i.e., \(S_W=1\). (e) The proposed weight updation (WU) unit with timing diagram.}
    \label{fig_SWC_and_WU_units} \vspace{-4mm}
\end{figure*}

Also, the methodology employed in ref.\cite{8345293} for signed dot product cannot be used for our purpose since it uses the change in voltage of BLB and BL line which are proportional to the decimal equivalent of weight and its \(1's\) complement, respectively, which cannot be added/subtracted directly in the analog domain for performing addition/subtraction operation. Therefore, a signed weight calculation (SWC) unit is designed, as shown in Fig.~\ref{fig_SWC_and_WU_units}. It works on the principle that one of the lines among BL (for negative weight) or BLB (for positive weight) will have voltage discharge proportional to the absolute magnitude of weight, as per from Eqs.~\ref{eqn_BL_line_discharge} and ~\ref{eqn_BLB_line_discharge}). As we have assumed earlier that signed weight stored in BCA will follow \(1's\) complement scheme for negative weight. This magnitude has to be converted to proportional negative or positive voltage. For this, first the sign \(S_W\) and \(V_{mux}\) are generated with the help of circuitry shown in Fig.~\ref{fig_SWC_and_WU_units}(a)), and corresponding expressions are:
\begin{equation} \label{eqn_Sign_Of_Weight}
    S_W=
    \begin{cases}
        0;    & \text{ for } V_{BLB}\geq V_{BL}, \text{ i.e., when } w \geq 0\\
        1;    & \text{ for } V_{BLB}<V_{BL}, \text{ i.e., when } w<0
    \end{cases}
\end{equation}
The output of MUX, \(V_{mux}\) is selected from among the BL and BLB since one of them will have absolute value of weight in terms of voltage change on the line (from Eqs.~\ref{eqn_BL_line_discharge} and ~\ref{eqn_BLB_line_discharge}). The selection is done using the sign of weights as the select line as:
\begin{equation} \label{eqn_Absolute_Weight}
    \left|w\right|=
    \begin{cases}
        \sum_{k=0}^{B_W-1}2^kb_k\propto \Delta V_{BLB}, \text{ if } S_W=0 \\
        \sum_{k=0}^{B_W-1}2^k\overline{b_k}\propto \Delta V_{BL}, \text{ if } S_W=1
    \end{cases}
\end{equation}

Thus, the output voltage of MUX, \(V_{mux}\) will contain the absolute value of weight as given by:
\begin{equation}
    V_{mux}=V_{PRE}-\Delta V_{lsb}\left|w\right|
\end{equation}
Thus the change in MUX voltage will be given by:
\begin{equation} \label{eqn_change_in_mux_voltage}
    \Delta V_{mux}=V_{PRE}-V_{mux}=\Delta V_{lsb}\left|w\right|\propto\left|w\right|
\end{equation}
\par
For calculation of signed weights the circuit in Fig.~\ref{fig_SWC_and_WU_units}(b) is employed. It uses OPAMP for maintaining the voltage polarity based on value of \(S_W\), can be seen from Fig.~\ref{fig_SWC_and_WU_units}(c)-(d)).  Thus, the output voltage at this stage will have positive or negative sign corresponding to positive or negative weights stored in the SRAM array and their magnitude will be proportional to the magnitude of corresponding weight, i.e., \(\left|w\right|\).

%These updated weights have to be accessed frequently during feedforward and backpropagation processes so an efficient hardware implementation is required to: \textit{B1)} avoid high energy and delay cost that will occur during frequent access of BCA for storing and accessing updated weight during each iteration of network training; and \textit{B2)} carry out addition in eqn. \ref{eqn_weight_update} for updating weights. Figure \ref{fig_SWC_and_WU_units}(e) shows the proposed hardware for weights updation during training. 

\begin{figure*}[t]
    \centering
    \includegraphics[width=175mm]{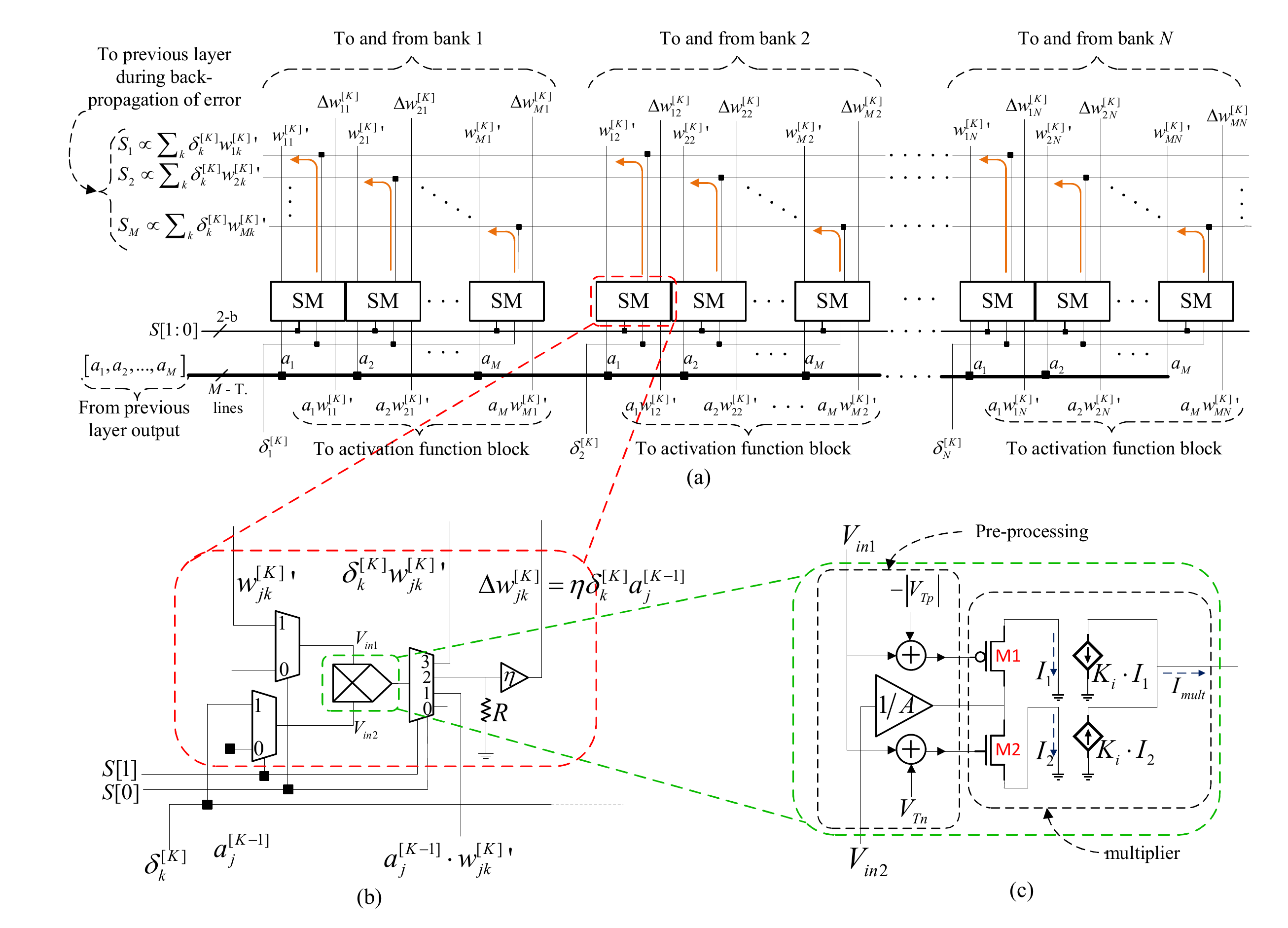}
    \caption{(a) Proposed hardware for carrying out backpropagation. (b) Realization of SM (Signed Multiplier) unit. (c) Four quadrant analog multiplier used inside SM unit.}
    \label{fig_Backpropagation} \vspace{-3mm}
\end{figure*}
\vspace{-3mm}
\subsection{Backpropagation and Weight Updation Units}
Backpropagation and weight updation are the most important part of any neural network for learning by adjusting its weight for accurate prediction and decision making until error is reduced to a minimum possible value. These updated weights have to be accessed frequently during feedforward and backpropagation processes that incur high energy and delay cost during frequent access of SRAM BCA for storing and accessing the updated weights during each iteration of network training. However, for improved throughput, energy efficiency, and on-chip training, we have designed a dedicated circuitry that eliminates this requirement and makes the proposed in-memory ANN architecture suitable for various AI/ML applications, as shown in Fig.~\ref{fig_SWC_and_WU_units}(e). In the proposed backpropagation and weight updation units, the generated signed weight \(w_{jk}^{[K]}\) is first sampled using switched connection \(\phi_S\) and stored in a sampling capacitor \(C_S\) shown in Fig.~\ref{fig_SWC_and_WU_units}(e)) instead of storing them in SRAM BCA. The \(C_S\) will be used to store all the intermediate weight corresponding to a single synaptic connection. The \(M\times N\) such weight updation units of Fig.~\ref{fig_SWC_and_WU_units}(e) will be required corresponding to the weights of all the \(M\times N\) connections involved in the \(K^{th}\) layer of the proposed architecture of the network, as shown in Fig.~\ref{fig_Single_Layer_Proposed}(a). Generalizing from Eqs.~\ref{eqn_Weight_update_output}-~\ref{eqn_weight_update_input}, the change in weight \(\Delta w_{jk}^{[K]}\) required for weight update is generated via SM unit (discussed in the next Sub-section) and is sampled on another sampling capacitor \(C_B\) via switched connection \(\phi_B\), shown in Fig.~\ref{fig_SWC_and_WU_units}(e)).

The unity gain buffer is used to access the voltage stored in the sampling capacitors. The charge leakage problem of capacitor will be self addressed since we are updating the capacitor charge at each iteration. Next, the outputs of each of these unity gain buffers are applied at the two ends of the series combination of two resistors of equal magnitude \(R\). Thus, the generated voltage at their common node \(V_U\) will be proportional to the sum of voltages corresponding to weight and change in weight, i.e., \(V_U=\left(w_{jk}^{[K]}+\Delta w_{jk}^{[K]}\right)/2\). The output is then latched/stored on another sampling capacitor \(C_L\) (via switched connection \(\phi_L\)) which will be used to update intermediate stage weights stored in sampling capacitor \(C_S\). But the voltage \(V_U\) have to be amplified by a factor of \(2\) to make the updated weight \(w_{jk}^{[K]}\)\('\) exactly equal to the sum of original weight and change in weight required. This amplification can be done using any suitable scheme. One of them may be to use an OPAMP based non-inverting amplifier of gain \(2\) as shown in Fig.~\ref{fig_SWC_and_WU_units}(e). The updated weight \(w_{jk}^{[K]}\)\('\) is then sent to the next unit for carrying out multiplication.

Weight update of the hidden layers is not a straight forward task as discussed in last section. Eqs.~\ref{eqn_weight_update_hidden} and ~\ref{eqn_weight_update_input} can be generalized to state that the weighted sum of the local gradient, i.e., \(\sum_k\delta_k^{[K]}w_{jk}^{[K]}\)\('\) is used by the \(j^{th}\) neuron of the previous layer for its weight update. The hardware implementation for the backpropagation block is shown in Fig.~\ref{fig_Backpropagation}(a). The product \(\delta_k^{[K]}w_{jk}^{[K]}\)\('\) obtained from the signed multiplier (SM) unit corresponding to \(j^{th}\) column of the \(k^{th}\) bank is added over all the \(k\)'s, i.e., over all banks to generate the sum \(S_j\) for \((j=1, 2,…, M\)), as shown in Fig.~\ref{fig_Backpropagation}(a). The final output at each backpropagating line will be proportional to the weighted sum of local gradient which will be fed to the previous layer for its weight update. Once the local gradient of all the layers have been calculated via the backpropagation process, then the weight update process can be parallized by just changing the control signal to \(S[1:0]=\)\textquoteleft\(10\)\textquoteright given in Table~\ref{Table_Control_Signal}, and discussed in detail in next Sub-section.

\begin{figure}[t]
    \centering
    \includegraphics[width=90mm]{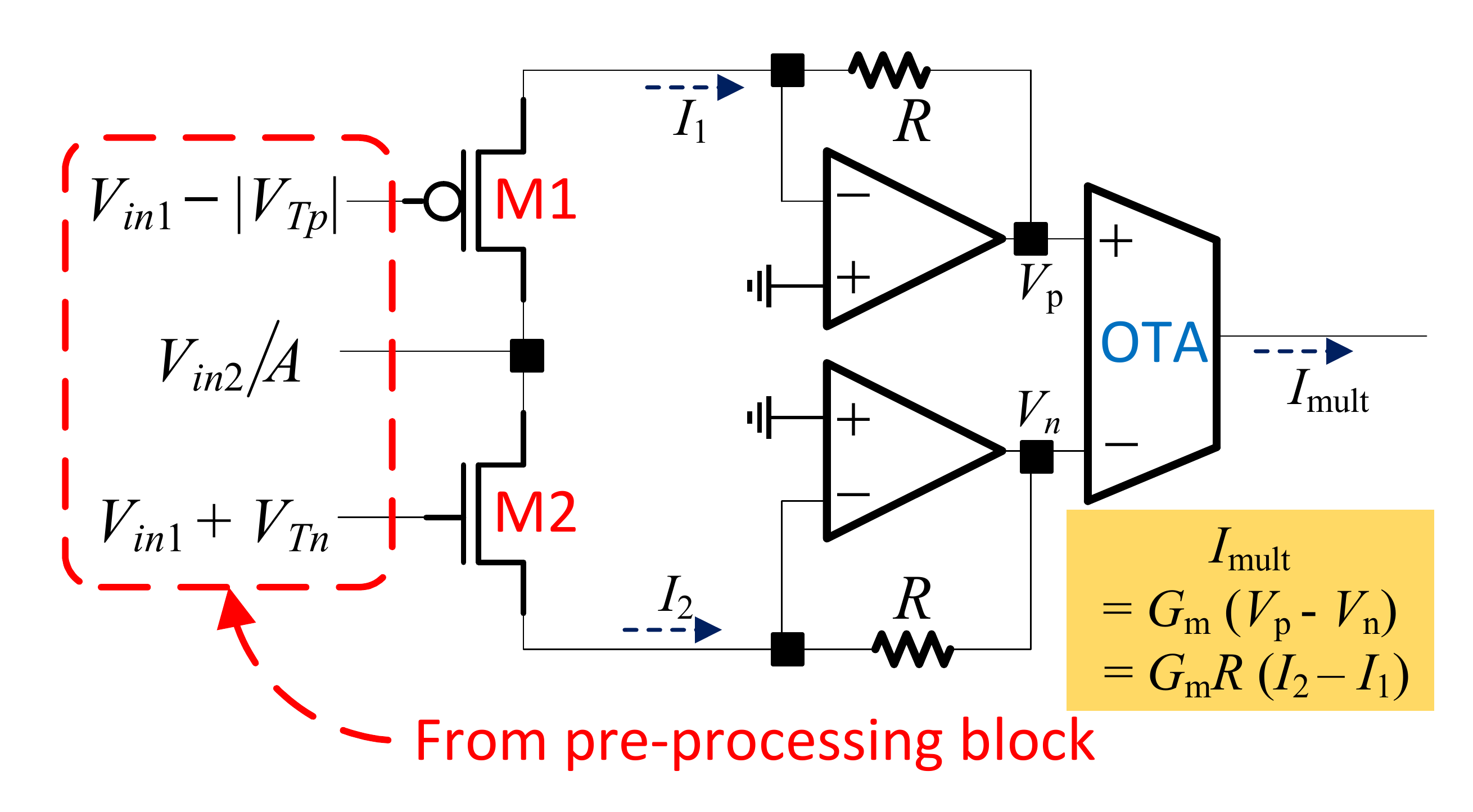}
    \caption{Realization of pair of current sources used in Fig.~\ref{fig_Backpropagation}(c).}
    \label{fig_Current_Source}\vspace{-5mm}
\end{figure}
\vspace{-3mm}
\subsection{Four Quadrant Analog Multiplier}

As discussed, different signals need to be multiplied during feedforward and backpropagation processes.
An accurate multiplier for the neural network is essential for better training and overall performance of the proposed in-memory ANN architecture. Fig.~\ref{fig_Backpropagation}(b) shows the architecture of the proposed signed multiplier (SM) unit employed to carry out multiplication of different signals depending upon the 2-b control signal \(S[1:0]\) applied to it. Depending upon control signals, the proposed multiplier will select different pairs of signals for multiplication, as shown in Table~\ref{Table_Control_Signal}. Fig.~\ref{fig_Backpropagation}(c) shows the hardware implementation of the four-quadrant analog multiplier which is the core of the proposed SM unit. In the proposed four-quadrant multiplier, the NMOS/PMOS work in the triode region to generate output proportional to the product of inputs \(V_{in1}\) and \(V_{in2}\). 

\begin{table}%[t]
\centering
    \begin{tabular}{|l|l|l|l|}
    \hline \rowcolor{lightgray}
        S[1] & S[0] & Product & Data Flow\\\hline
        0 & 0 & Forbidden & No where\\\hline
        0 & 1 & \(a_j^{[K-1]}w_{jk}^{[K]}\)\('\) & To activation function block\\\hline
        1 & 0 & \(\eta\delta_k^{[K]}a_j^{[K]}\) & To weight update unit\\\hline
        1 & 1 & \(\eta\delta_k^{[K]}w_{jk}^{[K]}\)\('\) & To backpropagation block\\
    \hline     
    \end{tabular}
    %\caption{Working of Signed Multiplier (SM) unit based on the applied control signal}
    \caption{Control signals for Signed Multiplier (SM) unit and corresponding operations/data flow.}
    \label{Table_Control_Signal} \vspace{-5mm}
\end{table}

%As already discussed, different signals need to be multiplied during feedforward and backpropagation. Further, the more accurate the multiplier output will be the more accurately the network will be trained and hence its performance will also be better. So we need an efficient (in terms of accuracy) hardware implementation of SM unit to that can be controlled to carry out multiplication among these signals during feedforward and backpropagation. Figure \ref{fig_Backpropagation}(b) shows the architecture of the proposed signed multiplier (SM) unit employed to carry out multiplication among different signals depending upon the 2-b control signal \(S[1:0]\) applied to it. Depending upon control signals, the proposed multiplier will select different pairs of signals for multiplication shown in Table \ref{Table_Control_Signal}. Figure \ref{fig_Backpropagation}(c) shows the hardware implementation of the four-quadrant analog multiplier which is the core of the proposed SM unit. The proposed four-quadrant multiplier works in the triode region to generate output proportional to the product of inputs \(V_{in1}\) and \(V_{in2}\). Its working is described below in detail.\par
For very small drain to source voltage, the drain current of NMOS in linear/triode region is given by \cite{sedra2004microelectronic}:
\begin{equation} \label{equ_NMOS_current_linear}
    i_{Dn}=\mu_n C_{ox} \left(\frac{W}{L}\right)\left(V_{GS}-V_{Tn}\right) V_{DS}
\end{equation}
where, \(\mu_n\) is the electron mobility, \(C_{ox}\) is the gate capacitance per unit area, \(\left(\frac{W}{L}\right)\) is the aspect ratio, \(V_{GS}\) and \(V_{DS}\) are gate to source and drain to source voltages, respectively, and \(V_{Tn}\) is the threshold voltage. Similarly, the drain current of PMOS is given by:
\begin{equation} \label{equ_PMOS_current_linear}
    i_{Dp}=\mu_p C_{ox} \left(\frac{W}{L}\right)\left(V_{SG}-\left|V_{Tp}\right|\right) V_{SD}
\end{equation}
where, \(\mu_p\) is the hole mobility, \(V_{Tp}\) is the threshold voltage of the PMOS, the rest of the quantities have the usual meaning as in Eq.~\ref{equ_NMOS_current_linear}. These Eqs.~\ref{equ_NMOS_current_linear} and ~\ref{equ_PMOS_current_linear} will be valid for two sets of conditions: (i) \(V_{GS}>V_{Tn}\) and \(V_{DS}\ll V_{GS}-V_{Tn}\) for NMOS and; (ii) \(V_{SG}>\left|V_{Tp}\right|\) and \(V_{SD}\ll V_{SG}-\left|V_{Tp}\right|\) for PMOS. To satisfy these conditions, the pre-processing block is designed, as shown in Fig.~\ref{fig_Backpropagation}(c)). Inside the pre-processing block, the voltage \(V_{in2}\) is reduced in amplitude by an amount \(A\) (reduction factor) that is sufficient enough to satisfy the necessary conditions on \(V_{DS}\) (for NMOS) and \(V_{SD}\) (for PMOS). There would be an obvious variation in multiplier output for different values of reduction factor, \(A\), which are discussed in the next Section along with simulations results. Next, if \(V_{GS}=V_{in1}+V_{Tn}\) (for NMOS) and \(V_{GS}=V_{in1}-\left|V_{Tp}\right|\) (for PMOS) is applied to the gate of the MOSFETs (as shown in Fig.~\ref{fig_Backpropagation}(c)) then, from Eqs.~\ref{equ_NMOS_current_linear} and ~\ref{equ_PMOS_current_linear}, the output current will be proportional to the product of input voltages. The above additions are performed inside the pre-processing block. Now, the outputs of the pre-processing block satisfies both necessary conditions ((i) and ii)) for the proposed multiplier to work properly. The outputs of the pre-processing block are fed to the multiplier, as shown in Fig.~\ref{fig_Backpropagation}(c). The analog multiplier generates current that is proportional to the product of inputs voltages \(V_{in1}\) and \(V_{in2}\). Two current sources of equal gain \(K_i\) are employed to generate the output current so that it could be used to drive the next stage. There are many ways to realize the pair of current sources used in Fig.~\ref{fig_Backpropagation}(c). One of them can be to use a pair of OPAMP and an OTA (Operational Transconductance Amplifier) as shown in Fig.~\ref{fig_Current_Source} which is used for simulations presented in next Section. The generated output current will be proportional to the product of input voltages within some offset error \(\in_m\), i.e., \(I_{mult}\propto \left(V_{in1}\cdot V_{in2}\pm\in_m\right)\). The occurrence of error will be due to small deviation from the linear behaviour of the MOSFETs in the triode region. This is because I-V characteristics of the MOSFET will not be perfectly linear in the triode region. But the effect of non-linearity can be reduced by decreasing the drain to source voltage which can be achieved by increasing the value of reduction factor \(A\) used in the pre-processing block in Fig.~\ref{fig_Backpropagation}(c). Thus, the effect of error, \(\in_m\), can be reduced by choosing a suitable value of the reduction factor \(A\). The variations of \(\in_m\) versus \(A\) are shown in next the Section. 
\vspace{-3mm}
\subsection{Activation Potential and Activation Function}
\begin{figure*}[t]
    \centering
    \includegraphics[width=168mm]{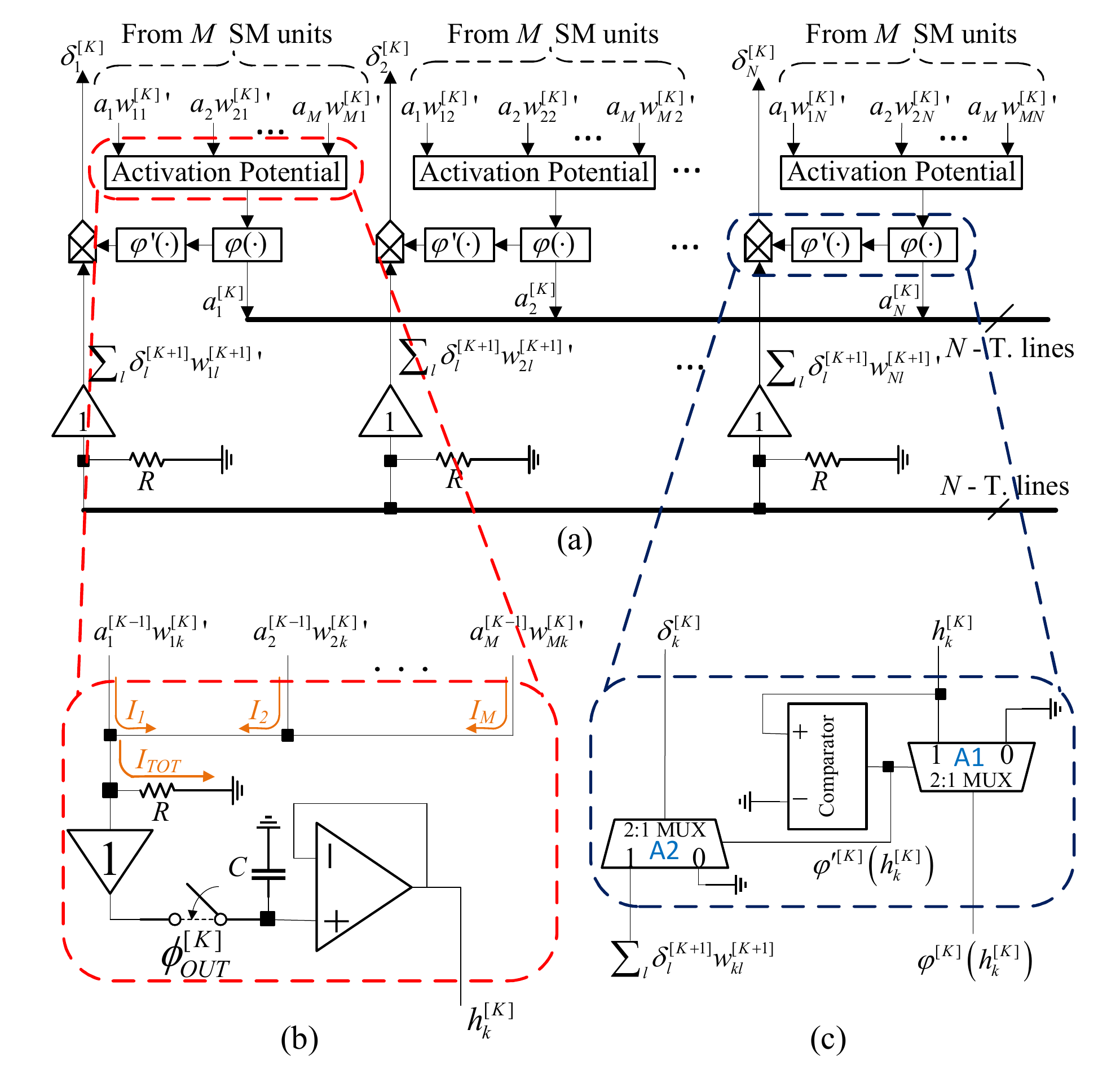}
    \caption{(a) Output stage at any layer of perceptron. (b) Summation of current to generate activation potential. (c) Hardware realization of ReLU activation function.}
    \label{fig_Output_of_layer_proposed} \vspace{-3mm}
\end{figure*}

Fig.~\ref{fig_Output_of_layer_proposed} shows the hardware implementation of activation potential and activation function for generating the output at any layer of perceptron. During feedforward, the product \(I_{mult}(\propto a_j^{[K-1]}w_{jk}^{[K]}\)\(')\) generated via SM unit and it is sent to the activation function block which carries out current based summation of input signals, as shown in Fig.~\ref{fig_Output_of_layer_proposed}(b). The total current \(I_{TOT}\) will be proportional to the weighted sum of inputs, i.e., \(I_{TOT}\propto\sum_ja_j^{[K-1]}w_{jk}^{[K]}\)\('\). But, we need output of the activation function in terms of voltage since the next layer requires input signal in the form of voltage instead of current. As input of multiplier requires signals in the form of voltage instead of current, a unity gain buffer is employed which converts the current through a resistor \(R\) to a voltage, as shown in Fig.~\ref{fig_Output_of_layer_proposed}(b). The value of \(R\) can be adjusted properly to make \(I_{TOT}\) exactly equal to the weighted sum of inputs. This activation potential is stored in a capacitor \(C\) and accessed via unity gain buffer. This is done to avoid its regeneration during backpropagation where it will be used for weight update. Activation functions perform a transformation on the input received, in order to keep values within a manageable range. There is a wide variety of activation functions available, such as linear, sigmoid, tanh, ReLU, etc. Out of these, the ReLU is the most popular and frequently used activation function for hidden layers. For the proposed in-memory ANN acrchitecture, we have employed and designed the ReLU activation function and its output can be expressed as:
% (\(R\) is chosen to be \(1\) k\(\Omega\) for simulation presented in next section).
\begin{equation} \label{eqn_Output_of_ReLU}
    a_k^{[K]}=\varphi\left(h_k^{[K]}\right)=max\left(0,h_k^{[K]}\right)
\end{equation}
where, \(h_k^{[K]}\) is the activation potential of the present layer and \(a_k^{[K]}\) is the output of the present layer generated after applying activation function \(\varphi(\cdot)\) on \(h_k^{[K]}\). Its differentiation is also very simple as given:
\begin{equation} \label{eqn_Differentiation_of_ReLU}
    \varphi '\left(h_k^{[K]}\right)=
    \begin{cases}
        0,&\text{ for } h_k^{[K]}\leq 0\\
        1,&\text{ for } h_k^{[K]} > 0
    \end{cases}
\end{equation}

Fig.~\ref{fig_Output_of_layer_proposed}(c) shows the hardware implementation of the ReLU function and its derivative. The MUX A1 performs the maximum operation as per Eq.~\ref{eqn_Output_of_ReLU} by selecting the maximum voltage based on comparator output. Another MUX A2 is used for generating the local gradient of this layer, which will be the product of differentiation of activation function and the weighted sum of the local gradient of the next layer, i.e., \(\delta_k^{[K]}=\left(\sum_l\delta_l^{[K+1]}w_{kl}^{[K+1]}\right)\varphi '\left(h_k^{[K+1]}\right)\). Using Eq.~\ref{eqn_Differentiation_of_ReLU} the local gradient of this layer can be expressed as:
\begin{equation} \label{eqn_local_gradient}
    \delta_k^{[K]}=
    \begin{cases}
        0,&\text{ for } h_k^{[K]}\leq 0\\
        \sum_l\delta_l^{[K+1]}w_{kl}^{[K+1]},&\text{ for } h_k^{[K]}>0
    \end{cases}
\end{equation}

The MUX A2 performs operation as per Eq.~\ref{eqn_local_gradient} based on the output of the comparator. But, ReLU is used only for the hidden layer. For output layer generally linear (for linear regression); sigmoid, tanh (for binary classification); or softmax (for multi-class classification) is used. The realization of linear activation function at the output is pretty much simple by just allowing the output to be equal to the input. Also, its derivative being equal to unity does not require any multiplier block for multiplying the error with derivative of activation function during backpropagation. But hardware realization of other non-linear activation function in the analog domain is quite complex. Instead, activation functions such as sigmoid\cite{sigmoid}, tanh\cite{tanh}, softmax\cite{softmax} and other non-linear activation functions can be easily realized in the digital domain using different approximation methods and/or look-up table (LUT) based methods. So, the incoming analog voltage corresponding to activation potential is first converted to the digital domain using ADC and then any of the above mentioned activation functions can be implemented digitally based on the requirement. In this paper, the softmax\cite{softmax} activation function have been employed and its output is converted back to the analog domain using DAC. 

\begin{figure*}[t]
    \centering
    \includegraphics[width=175mm]{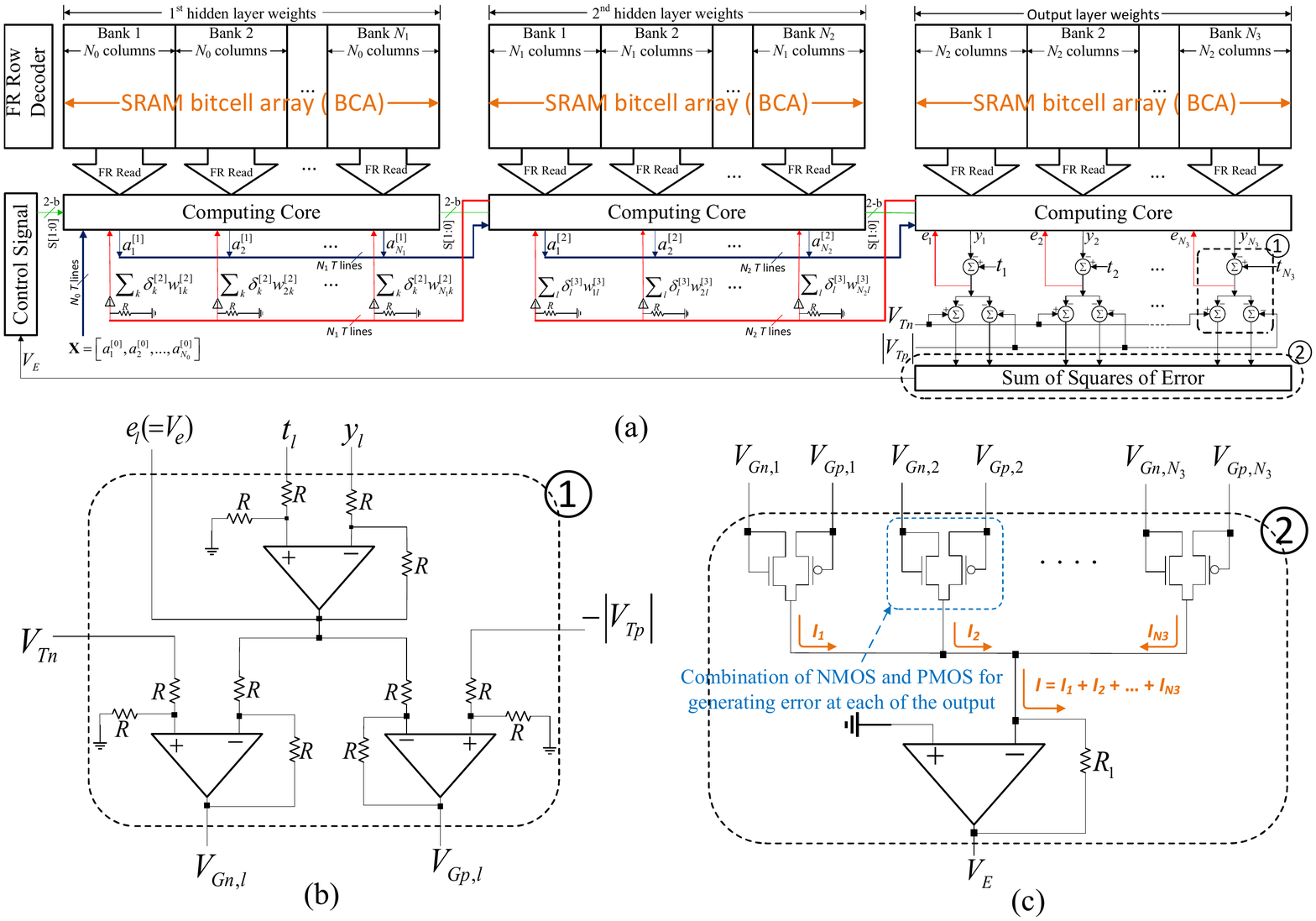}
    \caption{(a) Hardware realization of multilayered perceptron of Fig.~\ref{fig_Multilayered_Perceptron} by cascading multi-bank single layered architecture of fig.~\ref{fig_Single_Layer_Proposed}(a). (b) Generating error \(e_l\), and voltages- \(V_{Gn,l}\) and \(V_{Gp,l}\) via OPAM which is used in calculating sum of squares of error. (c) Generating sum of squares of error via MOSFETs at the output layer.}
    \label{fig_Multilayered_and_Error} \vspace{-3mm}
\end{figure*}
\vspace{-3mm}
\subsection{Multilayered Implementation Neural Network}
We have realized and discussed all the individual blocks needed for single layer of in-memory ANN architecture. However, for contemporary AI/ML applications, single layer neural network needs to be extended for multilayered neural network along with on-chip training capability. Fig.~\ref{fig_Multilayered_and_Error}(a) shows the multilayered perceptron model designed by cascading single-layered neural network. One of major advantages of the proposed neural network architecture is its scalability, as shown in Fig.~\ref{fig_Single_Layer_Proposed}(a), that is, any number of such architecture can be cascaded to design a multilayered perceptron with on-chip training capability. At the output layer, there will not be any further connections to the next layer but instead, there will be a mechanism for calculating the error of the network. The output of the sum of squares of error block is fed to the control signal block which controls the feedforward and backpropagation during the network training. The control signal block will stop the network training if the error \(V_E\) has stopped decreasing. The error at any neuron is given as the difference between target/intended output and the current output. For this, an OPAMP subtractor circuit is used, as shown in Fig.~\ref{fig_Multilayered_and_Error}(b) to calculate the difference \(e_l\left(=t_l-y_l\right)\) at the \(l^{th}\) neuron of the output layer. This difference is further sent to another two OPAMPs for calculating the difference \(V_{Gn,l}\left(=V_{Tn}-e_l\right)\) and \(V_{Gp,l}\left(=-\left|V_{Tp}\right|-e_l\right)\) which is used for calculating the sum of squares of error in next stage, where, \(V_{Gn,l}\) denotes the gate voltage of the NMOS at the \(l^{th}\) output, similarly \(V_{Gp,l}\) denotes the gate voltage of the PMOS at the \(l^{th}\) output node.

Consider the MOSFETs current in saturation region\cite{sedra2004microelectronic}:
\begin{equation} \label{eqn_MOSFET_current_saturation}
    I_D=
    \begin{cases}
        \frac{1}{2}k_n\left(V_{GS}-V_{Tn}\right)^2, & \text{ for NMOS}\\
        \frac{1}{2}k_p\left(V_{SG}-\left|V_{Tp}\right|\right)^2, & \text{ for PMOS}
    \end{cases}
\end{equation}
where, the parameters have usual meaning as mentioned in Eqs.~\ref{equ_NMOS_current_linear} and ~\ref{equ_PMOS_current_linear}. Now if \(V_{Gn,l}\left(=V_{Tn}-e_l\right)\), generated via Fig.~\ref{fig_Multilayered_and_Error}(b), is applied to the gate of NMOS then, for \(e_l<0\), the drain current of the NMOS would be:
\begin{equation} 
    I_{Dn,l}=\frac{1}{2}k_n\left(V_{Gn,l}-V_{Tn}\right)^2=\frac{1}{2}k_ne_l^2
\end{equation}

For \(e_l>0\), no current will flow through the NMOS which will wrongly reflect that the error of the network has become zero although it has actually not. So an innovative mechanism consisting of a combination of NMOS and PMOS having equal transconductance parameter \(k\left(=k_n=k_p\right)\) is designed, as shown in Fig.~\ref{fig_Multilayered_and_Error}(c) which will work even for \(e_l>0\). From Fig.~\ref{fig_Multilayered_and_Error}(c), the gate to source voltage for PMOS will be \(V_{Gp,l}\left(=-\left|V_{Tp}\right|-e_l\right)\) which will be negative and less than \(-\left|V_{Tp}\right|\) for \(e_l>0\), hence PMOS will be in ON state. At any instant, either the NMOS or PMOS will be in ON state at the \(l^{th}\) output, thus, the square of the error will be easily generated. Another important observation is that for the square of the error to be generated correctly, i.e., for Eq.~\ref{eqn_MOSFET_current_saturation} to be satisfied the MOSFET must always be in saturation which will happen when \(V_{DS}\geq V_{GS}-V_{Tn}\) (for NMOS) and \(V_{SD}\geq V_{SG}-\left|V_{Tp}\right|\) (for PMOS). For that, the same potential is applied to the drain and source, as shown in Fig.~\ref{fig_Multilayered_and_Error}(c). This will ensure that the condition for the MOSFET to be in saturation is always satisfied. Next is the generation of the sum of square of errors, for which an OPAMP is employed to sum up the current flowing through each of the MOSFET as:

\begin{equation} \label{eqn_Sum_of_Squares_of_Error}
    I=\sum_l\left(I_{Dn,l}+I_{Dp,l}\right)=\frac{1}{2}k\sum_le_l^2
\end{equation}

\begin{figure*}[t]
    \centering
    \subfigure[]{\label{fig_Flash_ADC}\includegraphics[width=0.49\textwidth]{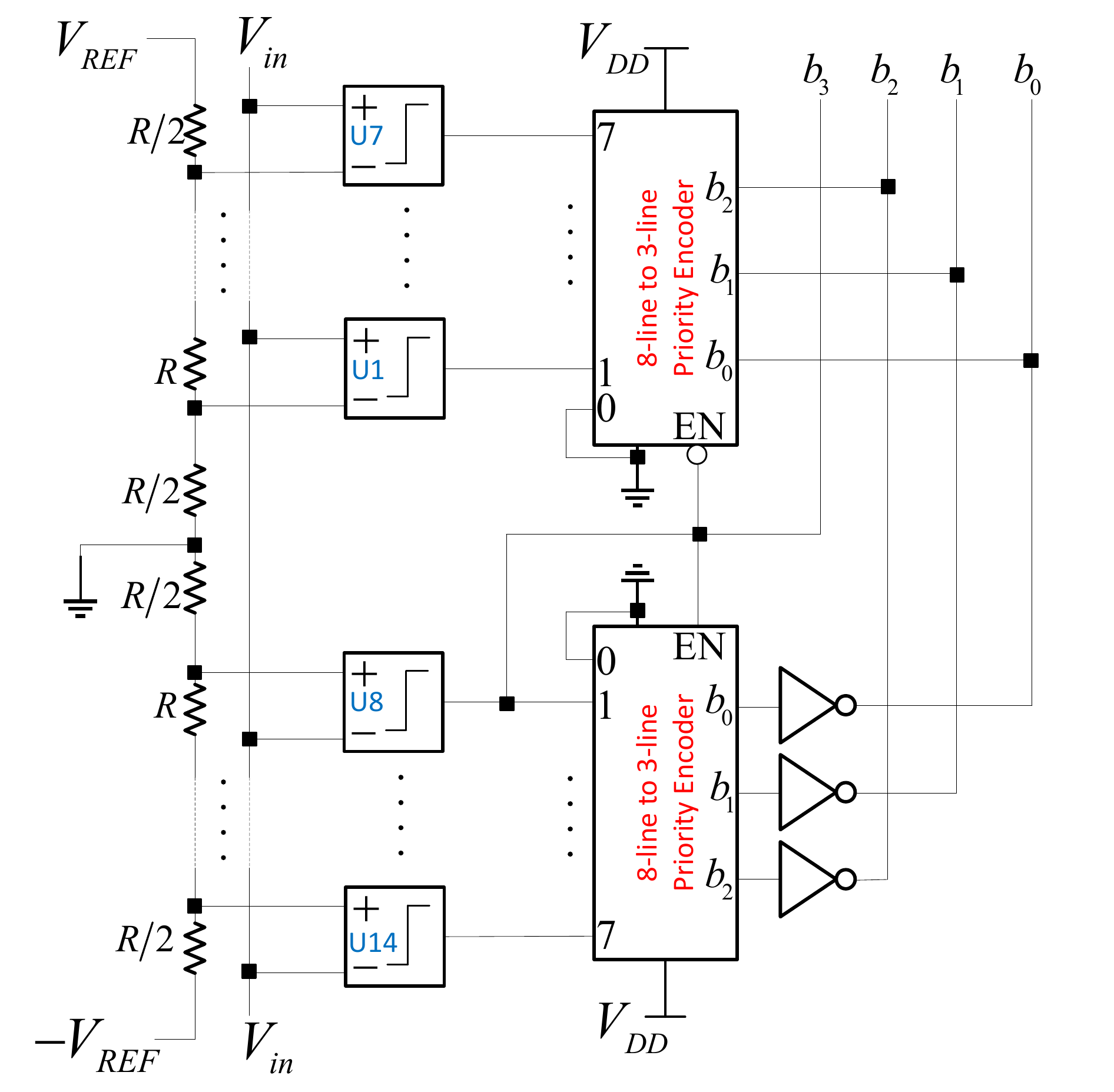}}
    \subfigure[]{\label{fig_Timing_Diagram}\includegraphics[width=0.49\textwidth]{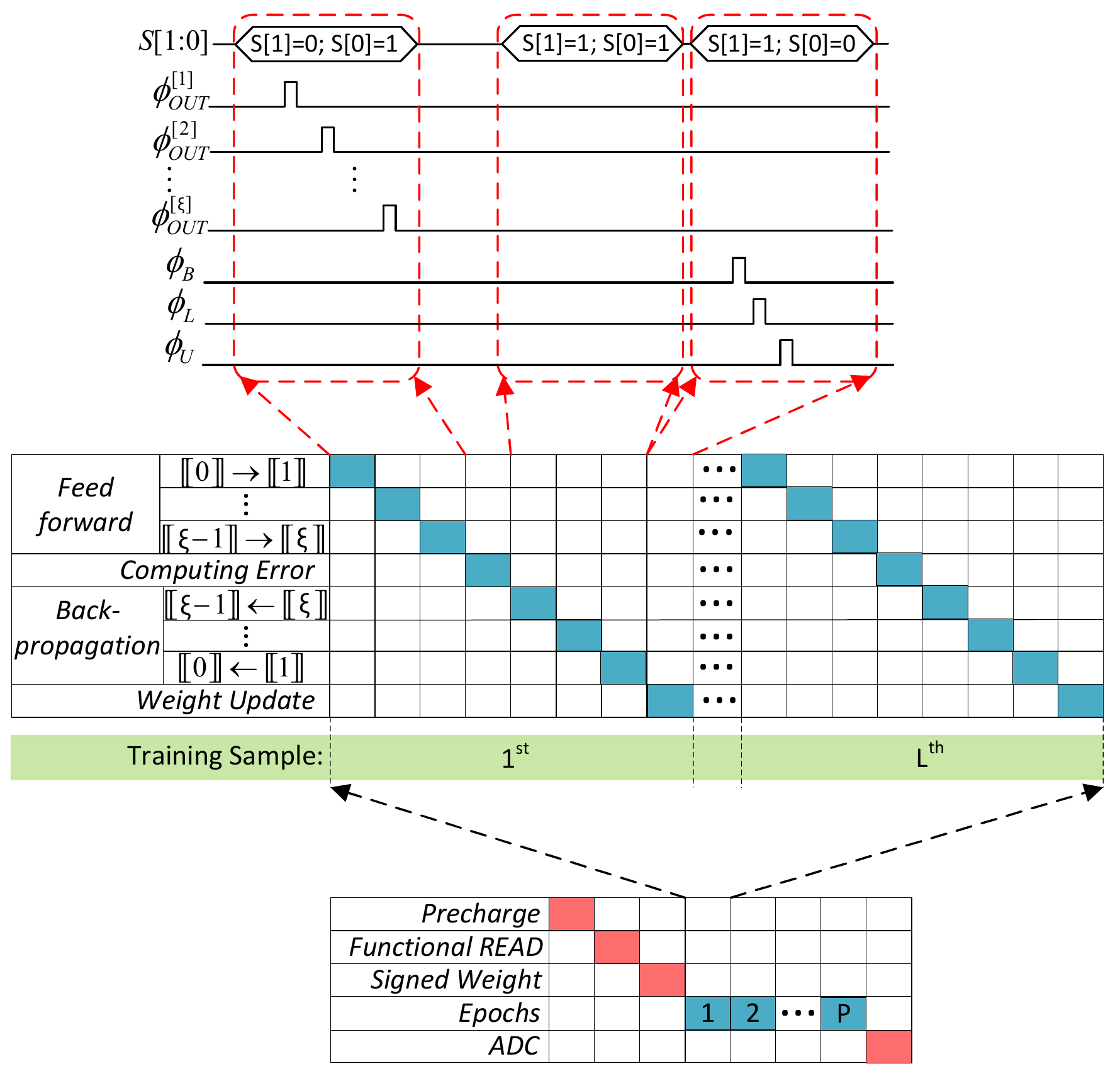}}
    \caption{(a) Signed Flash ADC design which converts negative weight using 1’s complement notation. (b) Timing diagram of the whole network.} \vspace{-3mm}
\end{figure*}

From above discussions, we can conclude for Eq.~\ref{eqn_Sum_of_Squares_of_Error} that either of the \(I_{Dn,l}\) or \(I_{Dp,l}\) will always be zero since the hardware implementation in Fig.~\ref{fig_Multilayered_and_Error}(c) is designed in such a way that only one of the MOSFETs (either NMOS or PMOS) will be ON at each of the output. Next, the final current flowing through OPAMP is converted to voltage \(V_E\) using resistor \(R_1\) as shown:
\begin{equation} \label{eqn_Sum_of_Square_of_Error_Voltage}
    V_E=-IR_1=-\frac{1}{2} k R_1 \sum_l e_l^2
\end{equation}
The negative sign is used since the voltage at the output of OPAMP will be negative when the current will flow in the conventional direction of NMOS as shown in Fig.~\ref{fig_Multilayered_and_Error}(c). Negative polarity of voltage can be ignored since we are only interested in the magnitude of the error and our aim will be to make the magnitude of error as small as possible.

\vspace{-3mm}
\subsection{Signed Flash ADC}
The trained weights have to be stored back in SRAM BCA to avoid training the network again and again. Since, the proposed architecture deals with negative/positive weights in the form of analog voltage so the general architecture of ADC needs to be modified to meet our requirements for the proposed in-memory ANN architecture. Specifically, for negative updated weights we need an ADC that uses \(1's\) complement rule for analog to digital conversion. Therefore, a 4-bit signed flash ADC is designed and presented in Fig.~\ref{fig_Flash_ADC}. The flash ADC is chosen due to its speed and optimal silicon chip area since we might need to convert a large number of updated weights (for very dense network) in the digital domain and store them back in SRAM BCA. In standard flash ADC, only positive reference voltage, i.e., \(V_{REF}\) are used as opposed to the proposed implementation where both negative and positive reference voltage \(V_{REF}\) and \(-V_{REF}\) have been used. Further, comparison of both positive and negative analog voltages (corresponding to proportional negative and positive updated weights) have been performed using resistor ladder network. The outputs of the comparators are fed to priority encoder for generating final output which is the digital equivalent of the magnitude of the input voltage \(V_{in}\). For negative \(V_{in}\), the desired output is \(1's\) complement of the digital equivalent of the magnitude of voltage for which NOT gate is used at the output of the priority encoder corresponding to negative reference voltage of \(-V_{REF}\) to generate its \(1's\) complement. 

For signed conversion using \(1's\) complement, the positive number have MSB equals \(0\) and negative number have MSB equals \(1\). To incorporate it, the MSB is chosen as the output of the first comparator (U8) employed for comparing negative voltage, as shown in Fig.~\ref{fig_Flash_ADC}). This will ensure the MSB \(b_3\) to be \(0\) for positive weights and \(1\) for negative weights lesser than \(-V_{res}/2\), where \(V_{res}\) is the resolution of the ADC in volts per step. Further, to reduce the output data line only one of the priority encoders is allowed to work at a time by applying \(b_3\) as enable signal to the EN input of each of the priority encoders. Thus, for \(b_3=0\) the priority encoder corresponding to positive reference voltage is enabled and for \(b_3=1\) the priority encoder corresponding to negative reference voltage is enabled. Hence, the designed ADC works for both positive and negative weights. The reference voltage \(V_{REF}\) decides the maximum swing of the ADC as well as the range of weights. The transfer function and value of the \(V_{REF}\) of the proposed signed flash ADC are discussed in detail in the next Section.

\begin{figure*}[t]
    \centering
    \subfigure[]{\label{sim_SWC}\includegraphics[width=0.45\textwidth]{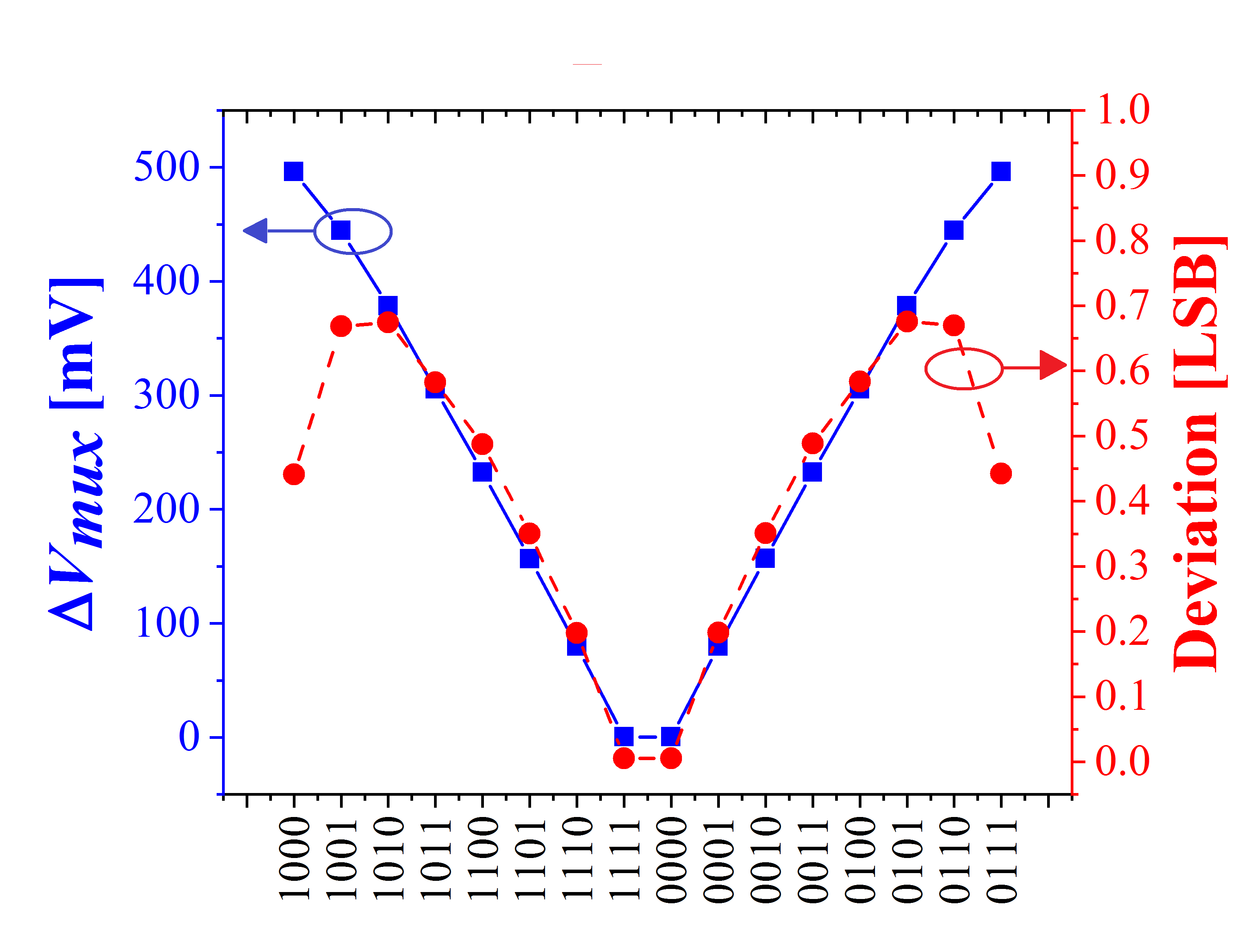}}
    \subfigure[]{\label{sim_FR_Energy}\includegraphics[width=0.45\textwidth]{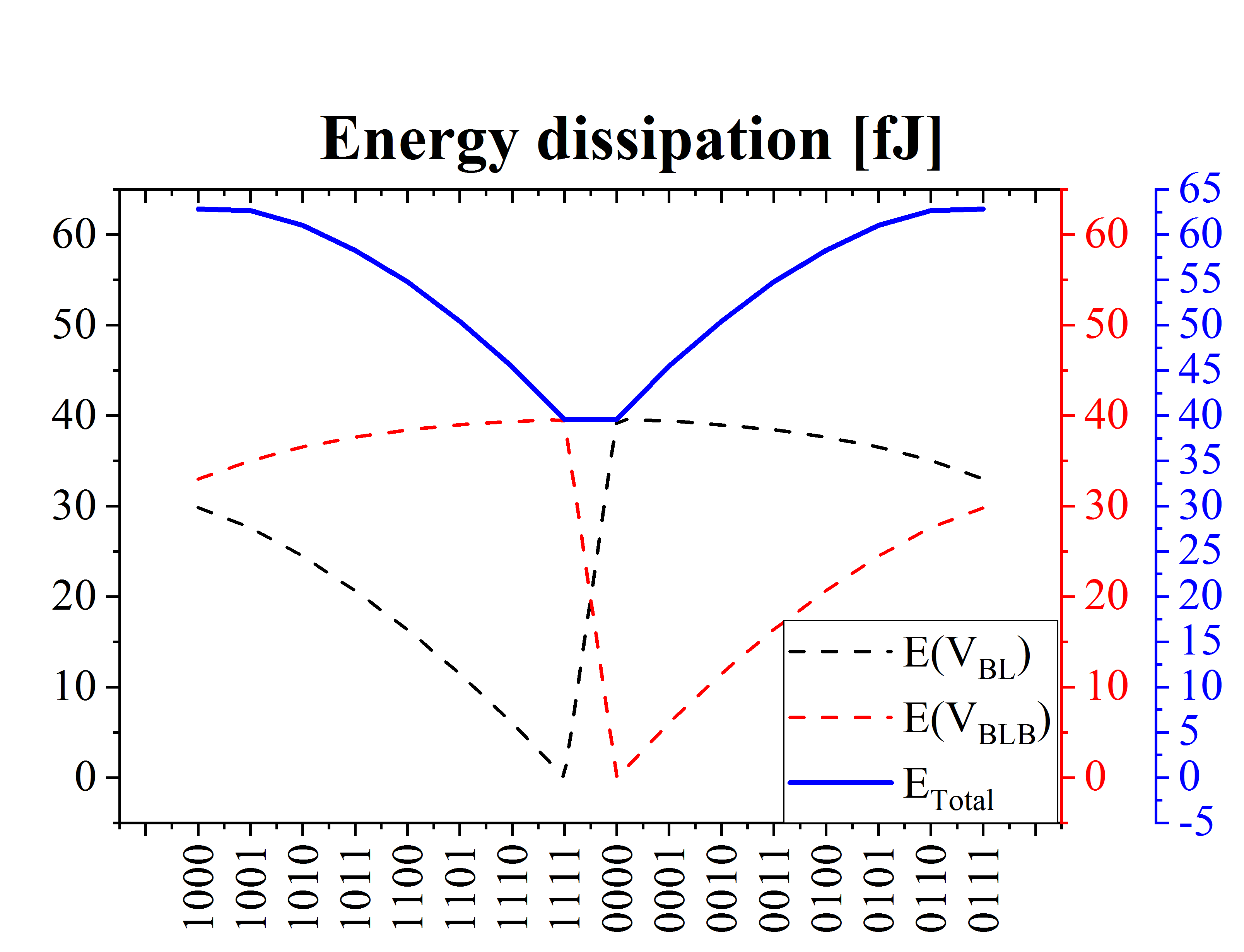}}
    \subfigure[]{\label{sim_WU}\includegraphics[width=0.45\textwidth]{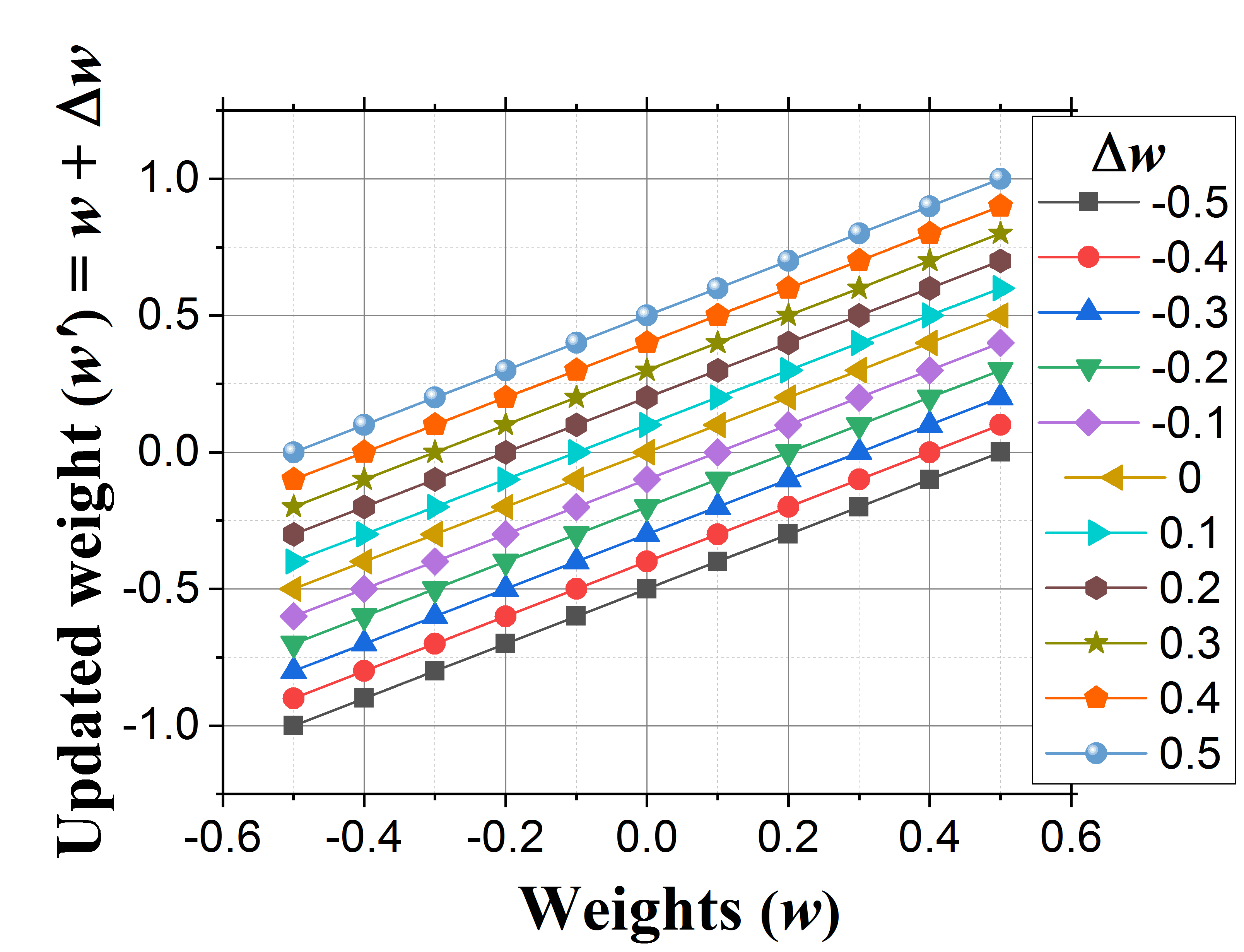}}
    \subfigure[]{\label{sim_Mult_Output}\includegraphics[width=0.45\textwidth]{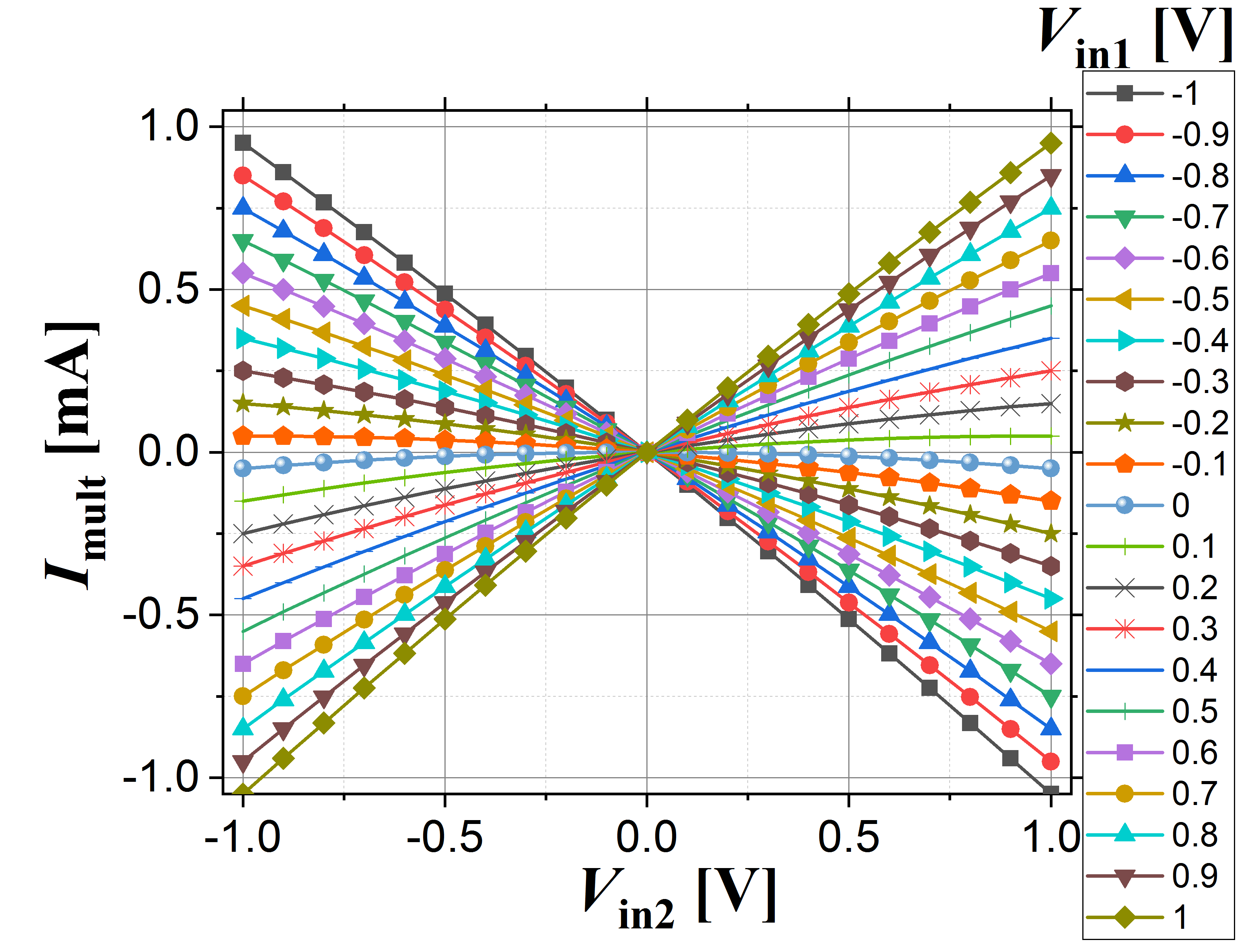}}
    \caption{(a) Simulation result of magnitude of weight calculated via \(\Delta V_{mux}\) inside SWC unit and its deviation (in LSB) from ideal (expected) output for 4-b resolution. (b) 4-b weight dependent energy dissipation during FR read process.(c) Simulation result of updated weights for \(\eta =1\). (d) Transfer function of the proposed multiplier of Fig.~\ref{fig_Backpropagation}(c).} \vspace{-3mm}
\end{figure*}

\vspace{-3mm}
\subsection{Timing Diagram}
Fig.~\ref{fig_Timing_Diagram} shows the timing diagram of the whole network. \textquoteleft{\fontfamily{lmss}\selectfont P}\textquoteright is the total number of epochs, \textquoteleft{\fontfamily{lmss}\selectfont L}\textquoteright is the total number of training samples presented to the network per epoch, \textquoteleft\(\xi\)\textquoteright is the total number of the layers. The notation: \([\![0]\!]\rightarrow[\![1]\!]\) denotes the forward propagation from input to the first hidden layer, and \([\![0]\!]\leftarrow[\![1]\!]\) denotes backpropagation from the first hidden layer to the input layer. With the same notation different index is used to indicate propagation between different layers. The working of the network starts with FR process of the SRAM BCA which fetches proportional analog equivalent of weights to the SWC units. Each SWC unit calculates the signed weight and sends the output to the WU unit. Once the FR process and SWC units have finish their work, then the whole training and testing procedure are controlled by \(2-b\) control signal \(S[1:0]\). During feedforward, the control signal switches to \textquoteleft01\textquoteright and calculates the activation potential at each layer which is latched to sampling capacitor, via switched connection \(\phi_{OUT}^{[K]}\), at the output of each layer. Next, the error of the network is calculated using the sum of squares of error cost function which is fed to the control block. If the error is still in reducing phase then the control block continues training the network otherwise the training is terminated. If the error has not stopped decreasing the next stage is the backpropagation phase (S[1:0]=\textquoteleft11\textquoteright) where the sum of local gradient is calculated inside the backpropagation block and is transmitted to previous layer via transmission lines. Once the backpropagation process completed then control signal switches to \textquoteleft10\textquoteright for weight updation. As training completes, the updated weights are stored back inside the SRAM BCA via signed flash ADC. Each of these major process is shown in timing diagram in Fig.~\ref{fig_Timing_Diagram}. Further, each of these major process is further divided into smaller sub-process for a clear insight into the working of the network.

\section{Simulation Results and Discussions}
In this Section, the simulation results and working of the proposed architecture are presented. Simulations of all the peripheral computing blocks were performed with SPICE simulator using High-Performance \(45\)nm PTM (Predictive Technology Model) models\cite{ptm}. The design parameter chosen for simulation are summerized in Table~\ref{Table_Simulation_Parameters}.

%In this Section, the simulation results and working of the proposed architecture are presented. All the simulations were performed on using standard six transistor SRAM cell/array and all the peripheral computing blocks are designed on LTSpice using High-Performance \(45\)nm PTM (Predictive Technology Model) technology\cite{ptm}. The design parameter chosen for simulation are summerized in Table \ref{Table_Simulation_Parameters}.

\begin{table}%[t]
\centering
    \begin{tabular}{|l|l|}
    \hline \rowcolor{lightgray}
        Parameter & Value \\\hline
        Technology & \(45\)nm PTM HP \\\hline
        \(V_{DD}\) & \(1\) V \\\hline
        \(V_{PRE}\) (of BL \& BLB) & \(1\) V \\\hline
        W/L & \(2\) \\\hline
        SRAM bit cell & \(6\)T \\\hline
        \(T_0\) (FR Read) & \(0.3\) ns \\\hline
        \(B_W\) & \(4\) \\\hline
        \(V_{REF}\) & \(0.496\) V \\
    \hline     
    \end{tabular}
    \caption{Design parameters for simulation} \vspace{-3mm}
    \label{Table_Simulation_Parameters}
\end{table}

\vspace{-3mm}
\subsection{Accuracy, Energy and Delay Analysis}

\begin{figure*}[t]
    \centering
    \subfigure[]{\label{sim_Mult_Energy}\includegraphics[width=0.45\textwidth]{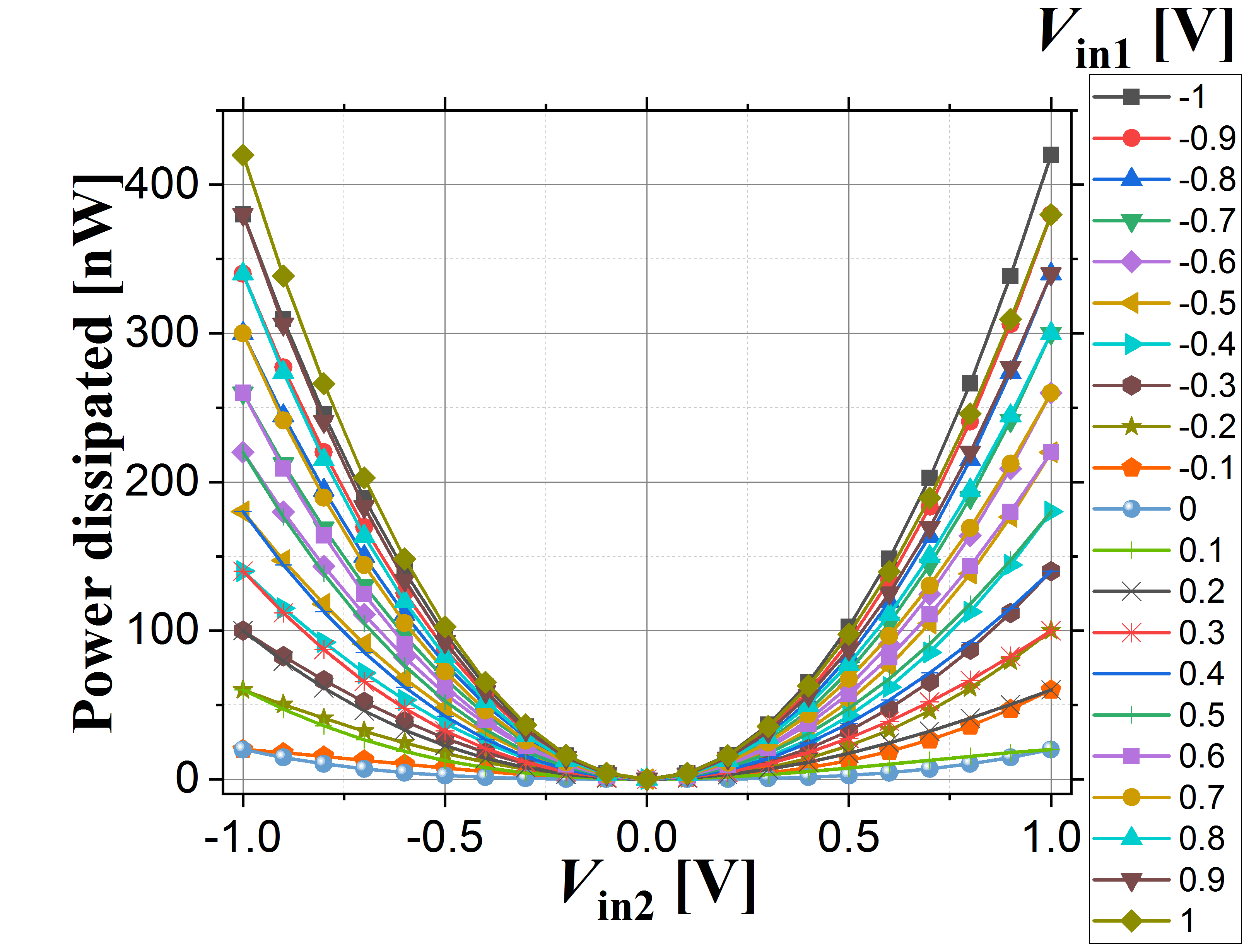}}
    \subfigure[]{\label{sim_Mult_Error}\includegraphics[width=0.45\textwidth]{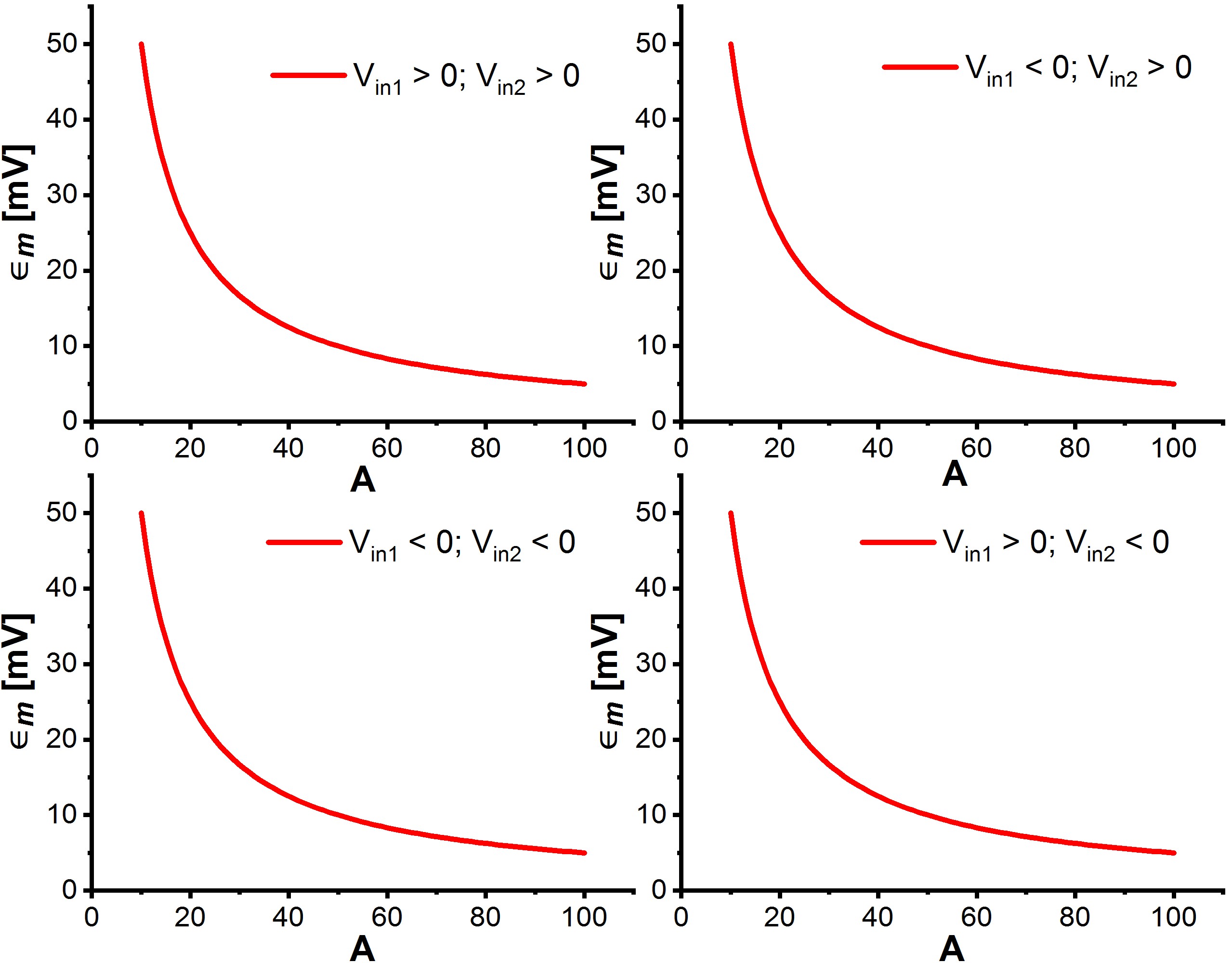}}
    \subfigure[]{\label{sim_ReLU}\includegraphics[width=0.45\textwidth]{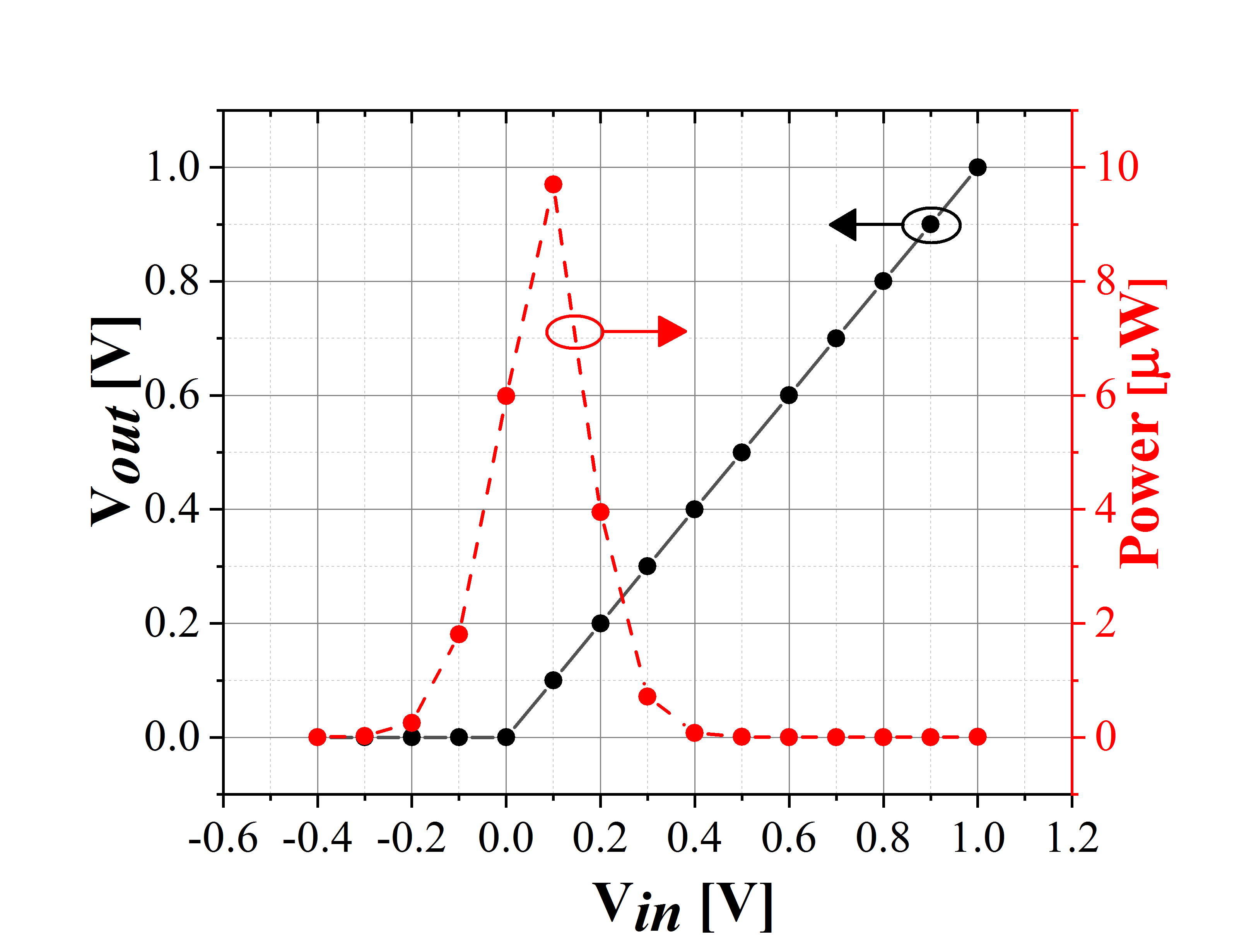}}
    \subfigure[]{\label{sim_Sum_Square_of_Error}\includegraphics[width=0.45\textwidth]{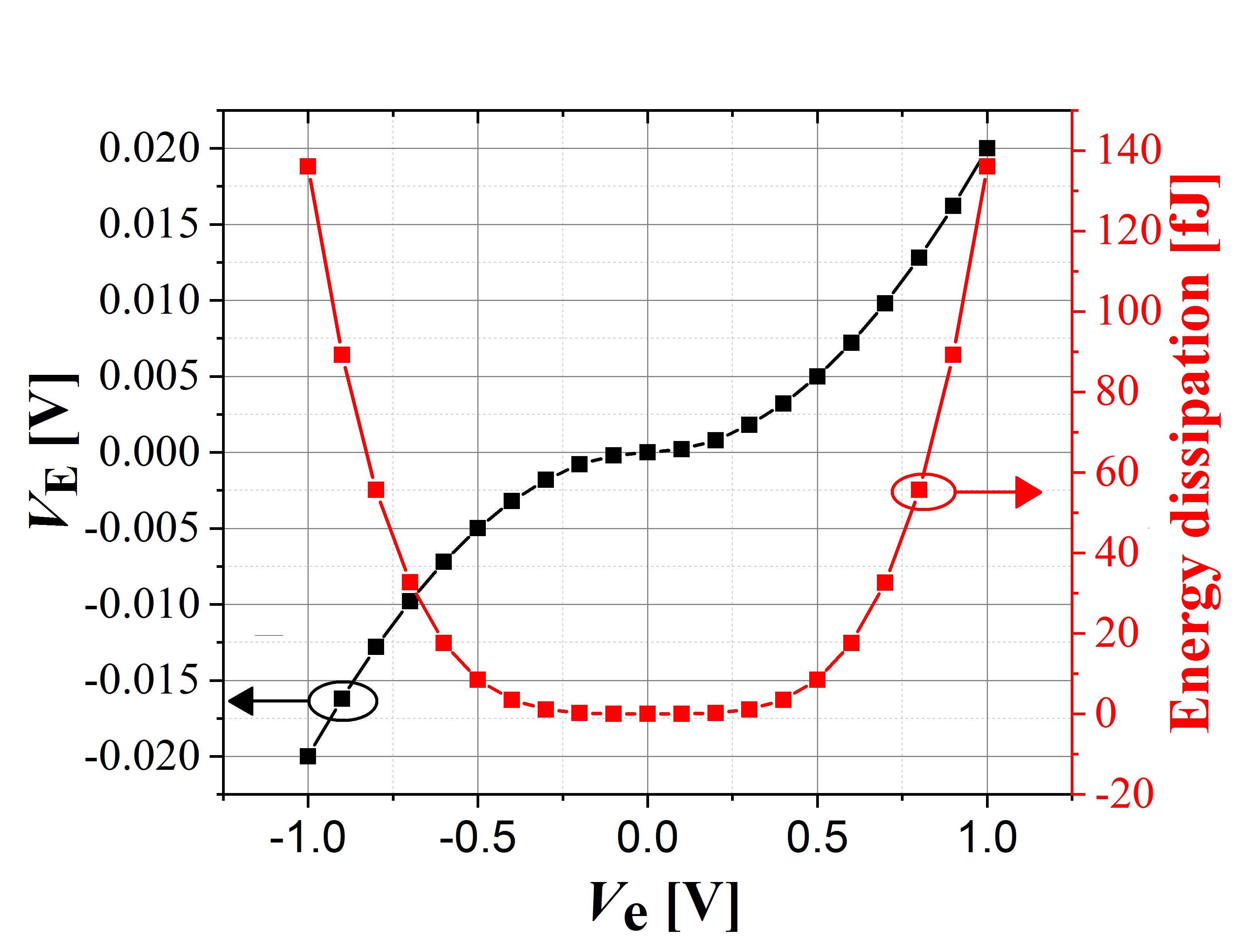}}
    \caption{(a) Energy dissipation in MOSFETs used inside multiplier of Fig.~\ref{fig_Backpropagation}(c). (b) Worst case analysis of error, \(\in_m\), associated with multiplier output with reduction factor \(A\) for different polarity of inputs \(V_{in1}\) and \(V_{in2}\) but each having equal magnitude of 1 V. (c) Transfer function and power dissipation of the proposed hardware for realizing ReLU activation function. (d) Simulation result of transfer function and energy dissipation of hardware (see Fig.~\ref{fig_Multilayered_and_Error}(c)) used for realizing square of error. The simulation results are for a single neuron at the output layer, i.e., for \(N_3=1\) in Fig.~\ref{fig_Multilayered_and_Error}(c).} \vspace{-5mm}
\end{figure*}

Accurate generation/updation of signed weights are essential for proper working of a perceptron network. Fig.~\ref{sim_SWC} shows the output of the magnitude of weight having 4-b resolution generated from SWC unit, as shown in Fig.~\ref{fig_SWC_and_WU_units}(a)-(d). It can be observed that the output is almost linear as expected (see Eq.~\ref{eqn_change_in_mux_voltage}) and the deviation of the simulated output is also within the expected range less than \(0.67\) LSB. A significant amount of energy dissipation is involved during the functional read (FR) process of the SRAM BCA. Fig.~\ref{sim_FR_Energy} shows the energy dissipation per column during the FR process of the SRAM BCA. From Fig.~\ref{sim_FR_Energy}, the energy expenditure is least when all the bits in the column-major format are either all zero or all one. As in this case either BL or BLB line will discharge completely and other will remain at a precharge voltage \(V_{PRE}\) level. However, with magnitude of weight increases both BL and BLB lines discharge by some amount, as a result, either the number of discharge paths, the discharge time, or both of these increases. Hence,  energy dissipation during the FR process increases with an increase in weight magnitude. The total energy dissipation during the FR process of the SRAM BCA can be modeled as:
%Figure\ref{sim_SWC} shows the output of the magnitude of weight having 4-b resolution generated via \(\Delta V_{mux}\) in fig. \ref{fig_SWC_and_WU_units}. The simulation result shows that the output is almost linear as it is expected to be from eqn. \ref{eqn_change_in_mux_voltage} and the deviation of the simulated output with the expected output was measured to be less than \(0.67\) LSB. A significant amount of energy dissipation is involved during the FR process of the BCA. Figure \ref{sim_FR_Energy} shows the energy dissipation per column during the FR process of the BCA. From fig. \ref{sim_FR_Energy}, the energy expenditure is least when either all the bits in the column are either all zero or all one. This is because in this case either BL or BLB line will discharge completely and other will remain at a precharge voltage of \(V_{PRE}\). But as the magnitude of weight increases, then: \textit{D1)} both of the BL and BLB line discharges by some amount, and \textit{D2)} either the number of discharge paths or the discharge time or both of these increases. Due to the reasons mentioned above in \textit{D1)} and \textit{D2)}, energy dissipation during the FR process increases with an increase in weight magnitude. The total energy dissipation during the FR process of BCA can be modeled as:
\begin{equation}
    E_{Total}(w)=\sum_k A_k(w) e^{-2\alpha k\cdot f(w)}
\end{equation}
where, \(\alpha=T_0/\tau_0\), \(T_0\) is pulse width applied at the WL of the LSB row of DIMA, \(\tau_0=C_{BL}R_{BL}\) is the time constant of the circuit, \(A_k(w)\) is the coefficient term that depends on the decimal equivalent of the weight stored in the SRAM BCA, and \(f(w)\) is a linear function of decimal equivalent of weight \((w)\). 

The maximum delay during FR read will be decided by the maximum time duration of the modulated pulse width WL signal applied to the MSB row which is \(2.4\) ns in our case for 4-b weights stored in SRAM BCA. Fig.~\ref{sim_WU} shows the variation of trained weight with signed weight (calculated in SWC unit) for different values of \(\Delta w\), for \(R=1\) k\(\Omega\) and learning rate \(\eta=1\). The variation is found to be linear from simulation results as it is expected from Eq.~\ref{eqn_weight_update}.

Fig.~\ref{sim_Mult_Output} shows the transfer function of the proposed multiplier (see Fig.~\ref{fig_Backpropagation}(c)). Designed multiplier employs a current controlled current source (CCCS) using a pair of OPAMP and an OTA, as shown in Fig.~\ref{fig_Current_Source} (as discussed in previous Section). The desired output is effectively achieved within some error, as can be seen from simulation results for a current gain factor \(K_i=250\). The value of \(R=25\) \(\Omega\) and \(G_m=10\) (for CCCS used in Fig.~\ref{fig_Current_Source}) was used to make the current gain \(K_i=G_m R\) equals \(250\) for simulation shown in Figs.~\ref{sim_Mult_Output} and ~\ref{sim_Mult_Energy}-\ref{sim_Mult_Error}. The error \(\in_m\) associated with the proposed multiplier can be reduced by increasing the reduction factor \(A\) used in the pre-processing block of Fig.~\ref{fig_Backpropagation}(c). For error analysis, multiplier output current was converted to a voltage by passing it through \(1\) k\(\Omega\) resistance and measuring the voltage across it. This is because for converting the multiplier output current into a voltage, wherever required, the same value of resistance of 1 k\(\Omega\) have been used throughout the simulations. From simulation results, it was observed that as the input voltage increases, the error \(\in_m\) associated with multiplier output also increases. Further, the error reaches its maximum when both the input voltage reaches 1 V each. Hence, for error analysis, both the inputs was kept at 1 V each for analysing worst case impact of error \(\in_m\) on the multiplier output. The error of output, \(\in_m\), was measured with respect to a reference voltage of 1 V which is the expected result for input combination of 1 V each. Fig.~\ref{sim_Mult_Error} shows the worst-case analysis of the error \(\in_m\) on multiplier output (for \(K_i=250\) and \(V_{in1}=V_{in2}=1\) V) versus the reduction factor \(A\). From Fig.~\ref{sim_Mult_Error}, it can be inferred that the proposed multiplier can be made more accurate by using even smaller value of drain to source voltage, i.e., by using even larger value of reduction factor, \(A\). But, with higher value of \(A\), we will need to use current amplifier with even more higher gain \(K_i\) than before to maintain output value of 1 V for input combinations of 1 V each. Hence, there will be a tradeoff between reduction factor \(A\) and the current gain factor, \(K_i\), of the multiplier. Fig.~\ref{sim_Mult_Energy} shows the power dissipation of the designed multiplier (See Fig.~\ref{fig_Backpropagation}(c)) as a function of applied input signals. Almost quadratic variation with input voltage is similar to that of a resistance because the MOSFETs in Fig.~\ref{fig_Backpropagation}(c) are designed to operate in linear region and behave as a voltage controlled resistance. 

Fig.~\ref{sim_ReLU} shows the transfer function and power dissipation of the proposed hardware for realizing ReLU activation function. The output of ReLU activation function is zero for inputs less than or equal to zero, and is equal to the input for inputs greater than zero as can be seen from Eq.~\ref{eqn_Output_of_ReLU}. The transfer function of the proposed ReLU block supports the Eq.~\ref{eqn_Output_of_ReLU} as can be seen from Fig.~\ref{sim_ReLU}. The power dissipation increases sharply when there is a transition in the activation potential which is in good agreement as maximum switching power dissipation takes place within the comparator. Further, the average delay incurred while generating output from the ReLU block was \(\approx 300\) ps. Another important parameter during the training is the total error of the network which have to keep track during weight updation process and it should be stopped as soon as the error turns out to be zero. Generally, this is very difficult to achieve, hence, a fixed number of epochs is chosen for training the network during which the error is continuously monitored till it reaches minimum after which training must be stopped. 

\begin{table}[t]
\centering
    \begin{tabular}{|l|l|}
    \hline \rowcolor{lightgray}
        Parameter & Value \\\hline
        \(N_{col,1}\)=No. of input layer neurons & 4 \\\hline
        No. of hidden layers & 1 \\\hline
        \(N_{bank,1}\)=No. of hidden layer neurons & 5 \\\hline
        \(N_{col,2}\)=No. of hidden layer neurons & 5 \\\hline
        \(N_{bank,2}\)=No. of output layer neurons & 3 \\\hline
        Loss function & Sum of squares of error \\\hline
        Optimizer & Gradient Descent \\\hline
        Learning rate (\(\eta\)) & 0.1 \\\hline
        \multirow{2}{*}{Activation Function} & ReLU (for hidden layer) \\
                                             & Softmax (for output layer) \\\hline
        Dataset & Iris\cite{Iris}  \\\hline
        \multicolumn{2}{|c|}{TRAINING} \\\hline
        Epochs & 500 \\\hline
        Accuracy & \(\approx 99\%\)\\\hline
        \multirow{2}{*}{Energy} & \(\approx 0.84\) \(\mu\)J/epoch \\
                                & (\(\approx 7.002\) nJ/iteration) \\\hline
        \multirow{2}{*}{Delay} & \(\approx 82.032\) \(\mu\)s/epoch \\
                                & \(\approx 0.683\) \(\mu\)s/iteration \\\hline
        \multicolumn{2}{|c|}{TESTING} \\\hline
        Accuracy & \(\approx 96.67\%\)\\\hline
        Energy/Decision(pJ) & \(1.855\)\\\hline
        Delay/Decision(ns) & \(680.6\)\\\hline
        Decision/s & \(1.47\) M\\\hline
        EDP/Decision(J\(\cdot\)s) & \(1.26\times 1e-18\)\\
    \hline     
    \end{tabular}
    \caption{Model description, parameters, and its training and testing results on Iris Dataset.} \vspace{-3mm}
    \label{Table_Model_Description}
\end{table}

Fig.~\ref{sim_Sum_Square_of_Error} shows the simulation results of the hardware designed for generating square of error at a single neuron at the output layer (for \(R_1=1\) k\(\Omega\)) which is seen to follow the square relation as per Eq.~\ref{eqn_square of error}. The total error will be the sum of the square of error at each of the outputs which will generate a voltage \(V_E\) at the output of OPAMP. The error voltage can be negative or positive due to the opposite direction of the current flowing through PMOS/NMOS devices. But, as stated earlier, we are interested in only to make the error zero or to make its magnitude as small as possible. The energy dissipation during this process will mainly be governed by the resistor \(R_1\) used in Fig.~\ref{fig_Multilayered_and_Error}(c). The simulation results in Fig.~\ref{sim_Sum_Square_of_Error} shows the energy dissipated for \(R_1=1\) k\(\Omega\). The time delay incurred during this process was calculated to be \(\approx 340\) ns. Fig.~\ref{sim_ADC} shows the output of the 4-b flash ADC designed for converting updated weights which corresponds to proportional analog voltages stored in sampling capacitor \(C_S\) inside WU units back to digital domain, and storing them back to SRAM BCA. The reference voltage \(V_{REF}\) was chosen to be \(0.496\) V which corresponds to a maximum binary output of 0111 during signed weight conversion (as shown in Fig.~\ref{sim_SWC}). The maximum INL (Integral Non-Linearity) of the ADC was observed to be around \(0.3V_{lsb}\), which is lower than the half of LSB.
\vspace{-3mm}
\subsection{Training and Testing on Iris Dataset}

\begin{figure*}[ht]
    \centering
    \subfigure[]{\label{sim_ADC}\includegraphics[width=0.45\textwidth]{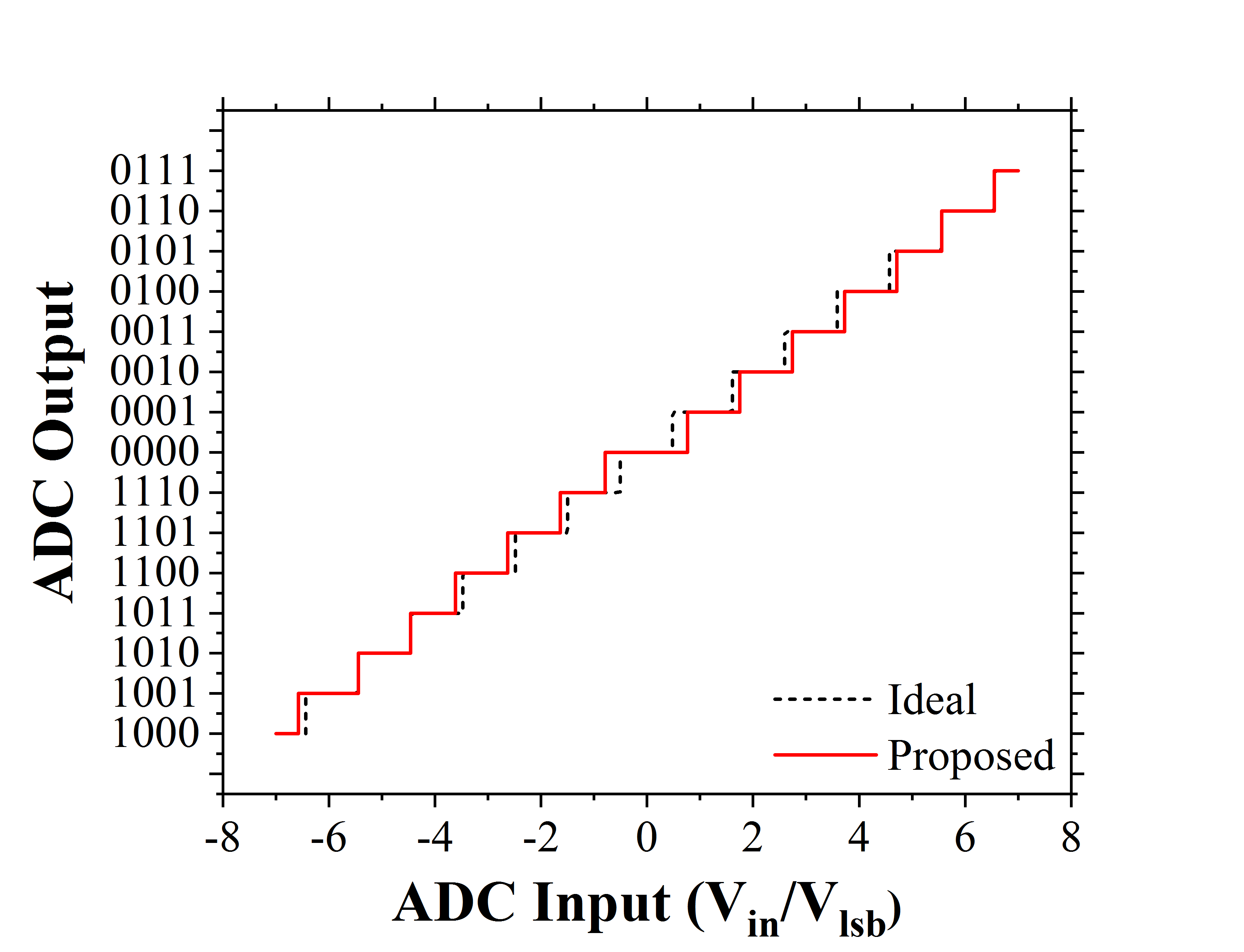}}
    \subfigure[]{\label{sim_Training}\includegraphics[width=0.45\textwidth]{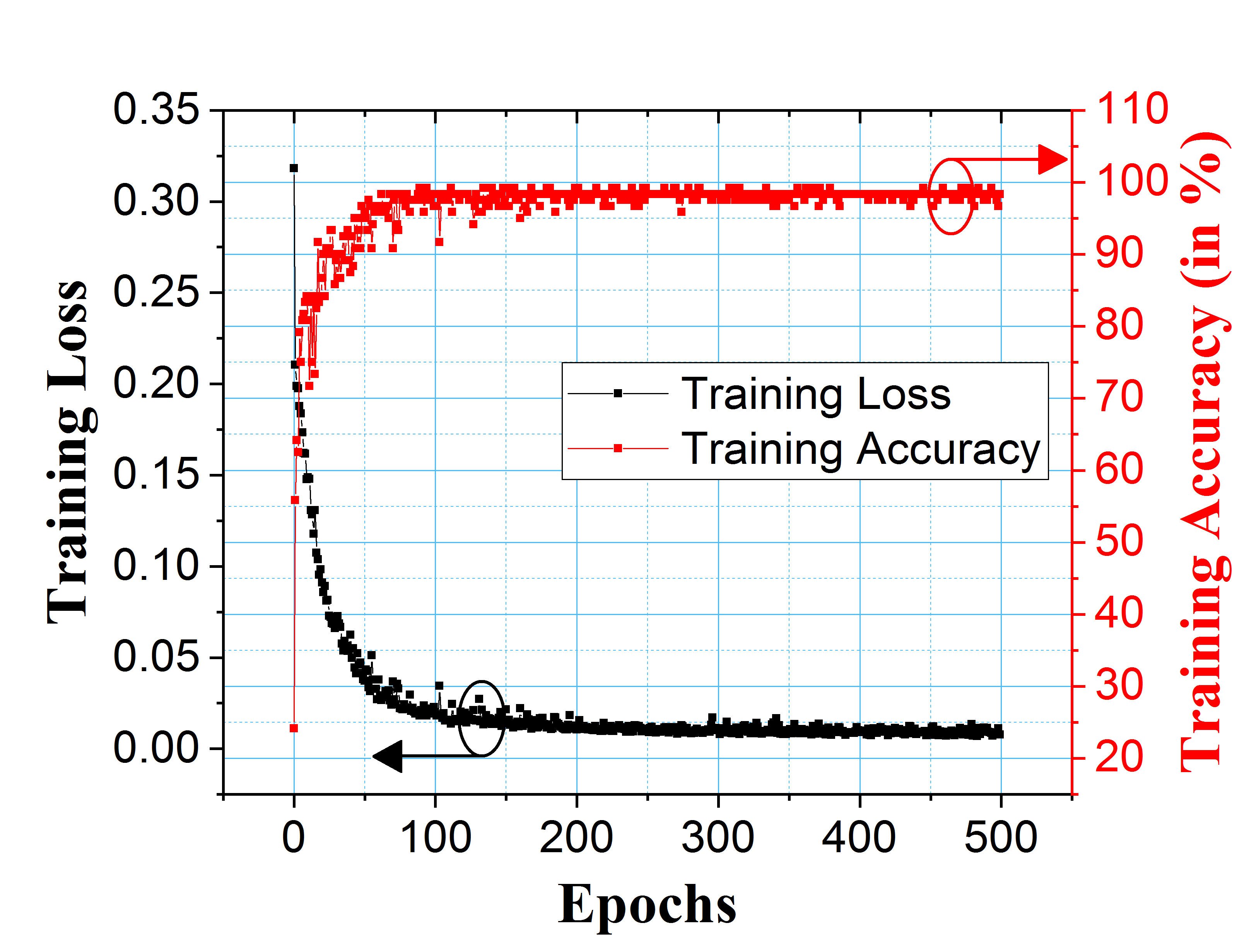}}
    \caption{(a) Output of the proposed signed flash ADC block. (b) Training loss and accuracy.} %\vspace{-3mm}
\end{figure*}

\begin{table*}[ht]
\centering
    \begin{tabular}{|l|l|l|l|l|l|l|}
    \hline \rowcolor{lightgray}
        . & Ref.\cite{7875410} & Ref.\cite{8246704} & Ref.\cite{8310397} & Ref.\cite{8310401} & Ref.\cite{8463601} & This work \\\hline
        Gate length (nm) & 130 & 65 & 65 & 65 & 65 & 45 \\\hline
        Technology & CMOS & CMOS & CMOS & CMOS & CMOS & PTM HP \\\hline
        Algorithm & Adaboost & SVM & CNN & DNN & SVM & ANN \\\hline
        Dataset & MNIST & MIT-CBCL & MNIST & MNIST & MIT-CBCL & Iris \\\hline
        Bitcell type & 6T & 6T & 10T & 6T & 6T & 6T \\\hline
        On-chip learning & No & No & No & No & Yes & Yes \\\hline
        Decision/s & 7.9 M & 9.2 M & - & - & 32 M & 1.47 M\\\hline
        \(B_W\) & 1 & 8 & 1 & 1 & 8 & 4\\\hline
        \(E_{MAC}^1\) (pJ) & 0.003 & 0.8 & 0.071 & 0.018 & 0.92 & 0.02 \\\hline
        Efficiency (TOPS/W) & 350 & 1.25 & 14 & 55.8 & 1.07 & 50 \\\hline
        EDP (J\(\cdot\)s) (\(\times\)1e-18) & 75.9 & 43.47 & - & - & 1.31 & 1.26 \\\hline
        \multicolumn{7}{l}{$^1$Energy of a single multiply-and-accumulate (MAC) operation.}
    \end{tabular}
    \caption{Comparison with other related works.} \vspace{-3mm}
    \label{Table_Comparison_with_Other_Work}
\end{table*}

The proposed in-memory ANN was trained and tested on Iris dataset~\cite{Iris} which consists of 150 records having 4 input features namely – petal length, petal width, sepal length, and sepal width. Each set of features corresponds to one of the species of Iris flower – Iris Setosa, Iris Versicolour, or Iris Virginica. Each of these three species have 50 records in the dataset. It is one of the most widely used dataset for training and testing the network for classification purpose. Since, it is a 3-class classification task so at the output layer softmax function was used which was implemented digitally as described in ref.\cite{softmax}. Table~\ref{Table_Model_Description} gives the sequentially trained model description, accuracy, energy and delay analysis on Iris dataset for 500 epochs. Table~\ref{Table_Comparison_with_Other_Work} compares the result of the proposed architecture with other earlier architectures. Fig.~\ref{sim_Training} shows the accuracy and loss value vs number of epochs. From Fig.~\ref{sim_Training}, the training accuracy reaches almost ~99\%. Further, the test accuracy was observed to be \(\approx 96.67\)\% and it can be further increased by: (a) incorporating the momentum term during the weight update which will reduce the possibility of the training mechanism to stuck in the local minima of the error landscape, and (b) using other cost function which works better for classification task, i.e., our proposed architecture will work fine for regression task with square of error cost function but for classification problem other cost functions such as cross entropy works even better. 

\vspace{-3mm}
\section{Conclusion}
In-memory on-chip trainable and scalable artificial neural network was designed and developed for wide variety of data intensive AI/ML algorithms. The main focus of the presented architecture was to exploit the in-memory analog computations for realizing the of ANN with on-chip training facility. The proposed architecture is scalable and re-configurable to map a large variety of AI/ML algorithms by enabling different activation functions and cost functions. The main strength of the proposed architecture lies in the fact that each steps from training to inference can be performed on-chip without external computing resources. Further, it does not require temporary registers/buffers for storing intermediate results that makes our proposed approach energy and computationally efficient. 

A neural network, being a complex interconnections of a large number of neurons, requires huge computations. Even some of the recently proposed architecture uses binary weights to reduce complexity but it may lead to accuracy issues since weights can have any real value (not necessarily only 0 and 1). Instead, in this paper, all the operations of a neural network are done in the analog domain which avoids such accuracy issues. However, some of the accuracy issues may arise during analog to digital conversion but that can be resolved using more number of bits/weight. Then the whole network was trained and tested on the iris dataset where the classification accuracy was estimated to be around \(\approx 96.67\) \%. Further, energy and delay analysis shows that the proposed work is \(\approx 46\times\) energy efficient in MAC (multiply-and-accumulate) operation as compared to previous work which employs DIMA. 
\vspace{-5mm}
\bibliography{sample}

\end{document}